\documentclass[usenatbib]{mnras}

\usepackage{epsfig}
\usepackage{newtxtext,newtxmath}
\usepackage[T1]{fontenc}
\usepackage{ae,aecompl}

\hypersetup{pdfauthor={E. O'Sullivan},
            pdftitle={Origin of the X-ray, radio and HI structures in the NGC 5903 galaxy group},
            pdfkeywords={galaxies: groups: individual (NGC 5903), galaxies: clusters: intracluster medium, galaxies: active, galaxies: interactions, X-ray: galaxies}}

\newcommand{\arcm}{\hbox{$^\prime$}}

\newcommand{\degree}{\hbox{$^\circ$}}
\newcommand{\rosat}{\emph{ROSAT}}
\newcommand{\chandra}{\emph{Chandra}}
\newcommand{\xmm}{\emph{XMM-Newton}}
\newcommand{\xmms}{\emph{XMM}}

\newcommand{\arcs}{\mbox{\arcm\arcm}}

\newcommand{\Zsol}{\ensuremath{Z_{\odot}}}
\newcommand{\Lsol}{\ensuremath{L_{\odot}}}
\newcommand{\Msol}{\ensuremath{M_{\odot}}}
\newcommand{\Msolpyr}{\ensuremath{M_{\odot}~yr^{-1}}}

\newcommand{\s}{\ensuremath{\mbox{~s}}}
\newcommand{\ps}{\ensuremath{\s^{-1}}}
\newcommand{\cm}{\ensuremath{\mbox{~cm}}}
\newcommand{\pcmsq}{\ensuremath{\cm^{-2}}}
\newcommand{\km}{\ensuremath{\mbox{~km}}}

\newcommand{\erg}{\ensuremath{\mbox{~erg}}}
\newcommand{\ergps}{\ensuremath{\erg \ps}}
\newcommand{\ergpspcmsq}{\ensuremath{\erg \ps \pcmsq}}
\newcommand{\kmps}{\ensuremath{\km \ps}}

\newcommand{\Dtf}{\ensuremath{D_{\mathrm{25}}}}

\newcommand{\gtsim}{\,\rlap{\raise 0.5ex\hbox{$>$}}{\lower 1.0ex\hbox{$\sim$}}\,} 

\newcommand{\Hi}{H\textsc{i}}

\newcommand{\nh}{\ensuremath{\mathrm{n}_\mathrm{H}}}

\begin{document}

\title[ 
X--ray, radio and HI structures in NGC~5903
] 
{ 
The origin of the X--ray, radio and HI structures in the NGC~5903 galaxy group\footnotemark[1]
}

\author[E. O'Sullivan et al.]  {Ewan O'Sullivan$^{1}$\footnotemark[2], Konstantinos Kolokythas$^{2}$, Nimisha G. Kantharia$^{3}$, 
\newauthor Somak Raychaudhury$^{2,4}$, Laurence P. David$^{1}$, Jan M. Vrtilek$^{1}$\\
  $^{1}$ Harvard-Smithsonian Center for Astrophysics, 60 Garden Street, Cambridge, MA 02138, USA\\
  $^{2}$ Inter-University Centre for Astronomy and Astrophysics, Pune 411007, India\\
  $^{3}$ National Center for Radio Astrophysics/TIFR, Pune University Campus, Pune 411 007, India\\
  $^{4}$ Department of Physics, Presidency University, 86/1 College Street, Kolkata 700073, India}

\date{Accepted 2017 October 12; Received 2017 October 12; in original form 2017 June 05}

\pagerange{\pageref{firstpage}--\pageref{lastpage}} \pubyear{2010}

\maketitle

\label{firstpage}

\begin{abstract} 
The NGC~5903 galaxy group is a nearby ($\sim$30~Mpc) system of $\sim$30 members, dominated by the giant ellipticals NGC~5903 and NGC~5898. The group contains two unusual structures, a $\sim$110~kpc long \Hi\ filament crossing NGC~5903, and a $\sim$75~kpc wide diffuse, steep-spectrum radio source of unknown origin which overlaps NGC~5903 and appears to be partly enclosed by the \Hi\ filament. Using a combination of \chandra, \xmm, GMRT and VLA observations, we detect a previously unknown $\sim$0.65~keV intra-group medium filling the volume within 145~kpc of NGC~5903, and find a loop of enhanced X-ray emission extending $\sim$35~kpc southwest from the galaxy, enclosing the brightest part of the radio source. The northern and eastern parts of this X-ray structure are also strongly correlated with the southern parts of the \Hi\ filament. We determine the spectral index of the bright radio emission to be $\alpha_{150}^{612}$=1.03$\pm$0.08, indicating a radiative age $>$360~Myr. We discuss the origin of the correlated radio, X-ray and \Hi\ structures, either through an interaction-triggered AGN outburst with enthalpy 1.8$\times$10$^{57}$~erg, or via a high-velocity collision between a galaxy and the \Hi\ filament. While neither scenario provides a complete explanation, we find that an AGN outburst is the most likely source of the principal X-ray and radio structures. However, it is clear that galaxy interactions continue to play an important role in the development of this relatively highly evolved galaxy group. We also resolve the question of whether the group member galaxy ESO~514-3 hosts a double-lobed radio source, confirming that the source is a superposed background AGN.
\end{abstract}


\footnotetext[1]{Based on observations obtained with XMM-Newton, an ESA science mission with instruments and contributions directly funded by ESA Member States and NASA.}
\footnotetext[2]{E-mail: eosullivan@cfa.harvard.edu}

\section{Introduction}
\label{sec:intro}
Galaxy groups are a diverse population, ranging from spiral-rich systems with large reservoirs of atomic and molecular gas located in and around the galaxy discs, to elliptical-dominated systems whose dominant gaseous component is the X-ray luminous plasma of the hot intra-group medium (IGM). This diversity is widely considered to be the product of dynamical evolution and mass assembly. As groups grow, their galaxy population evolves through merger, while accretion and mergers with other groups increases the depth of their gravitational potential to the point where infalling gas is gravitationally shock heated into the X-ray emitting regime.

Many authors have used compact groups as laboratories to study the process of group evolution. \citet{VerdesMontenegroetal01} propose a model in which tidal interactions between the spiral galaxies of dynamically young groups remove much of the cold neutral gas from their discs, producing intergalactic clouds or filaments, or even a diffuse cold intra-group halo. \citet{Konstantopoulosetal10} expand on this model and argue that groups can follow two evolutionary paths. In systems whose evolution is slow, star formation exhausts much of the gas in galaxies before they interact, leading to ``dry'' mergers and little development of the IGM. In groups where interactions occur early in development, galaxy mergers are gas-rich (``wet'' mergers), producing starbursts and outflowing winds which, along with stripped cold gas, contribute to the formation of the hot, X-ray luminous IGM. 

The role of cold gas in group evolution is not well understood. In part, this is because we have few examples of groups in an intermediate evolutionary phase, where a hot IGM has begun to form but a significant amount of cold gas is still present. Spiral-rich groups are known to be less X-ray luminous than their elliptical dominated counterparts \citep{Mulchaeyetal03}, with emission associated with the galaxies, not the IGM \citep{OsmondPonman04}. Among the few cases where cold gas coexists with a hot IGM are HCG~16 \citep{OSullivanetal14c,Belsoleetal03,Ponmanetal96} and Stephan's Quintet, HCG~92 \citep{OSullivanetal09,Trinchierietal03}. The latter is of particular interest since it contains a unique structure; an intergalactic shock front where an infalling galaxy has collided with a filament of neutral hydrogen, shock heating a $\sim$40~kpc long section into X-ray and radio-luminous plasma \citep{Sulenticetal01,Xuetal03}. This may be an example of the process by which cold gas stripped from galaxies contributes to the build-up of the hot IGM. 
 
More examples of this kind of shock are required if we are to understand
this process and determine its importance to group evolution.
Unfortunately, such shocks have a limited period of visibility \citep[a few
tens of Myr,][]{VanderhulstRots81}, and they are expected in relatively
X-ray faint groups, making them difficult to find in current X-ray and
radio surveys.

In more evolved groups where the hot IGM is fully established, AGN play a key role in its thermal regulation. It has been shown that in many X-ray bright groups, cavities formed by the inflation of radio lobes from the central galaxy contain sufficient energy to balance radiative cooling \citep[e.g.,][]{Birzanetal08,Cavagnoloetal10,OSullivanetal11b}. In general, these group-central AGN are thought to function as part of a feedback loop, fuelled by cooling from the IGM. However, any flow of gas into the centre of a galaxy hosting a super-massive black hole can in principle fuel an AGN, and cold-gas rich mergers are thought to trigger powerful outbursts in some systems \citep{Tadhunteretal16}. This raises the question of whether the galaxy interactions expected in the earlier stages of group evolution can trigger nuclear activity, either through gas-rich mergers or infall of stripped gas into ellipticals. Examination of \Hi-rich ellipticals suggests that they typically host low-power, compact sources rather than jet systems \citep[e.g.,][]{Emontsetal10}. There are counter-examples, e.g., NGC~1167, which possesses an extended \Hi\ disc \citep{Struveetal10}, a currently active compact FR-I source, but also a pair of old radio lobes extending $>$100~kpc \citep{Shulevskietal12,Brienzaetal16}. However, the connection between the \Hi\ and radio activity in these cases is often unclear.

In this paper we use a combination of X-ray and radio continuum observations to examine the NGC~5903 group, which shows similarities with both
Stephan's Quintet, and more evolved systems whose IGM is regulated by AGN feedback. We describe previous studies of the group in
Section~\ref{sec:N5903} and our observations and data reduction in
Section~\ref{sec:obs}. We outline the results of our GMRT and VLA
multi-frequency radio continuum analysis in Section~\ref{sec:radio_ims},
and \chandra\ and \xmm\
X-ray analysis in Section~\ref{sec:xray}. We discuss these results in
Section~\ref{sec:disc} and present our conclusions in
Section~\ref{sec:conc}. Throughout the paper, we adopt the redshift of
NGC~5903 \citep[$z$=0.008556,][]{Smithetal00} for the group as a whole and
a surface brightness fluctuation (SBF) distance of 31.5~Mpc from
\citet{Tonryetal01}, corrected to match the Cepheid zero--point of
\citet{Freedmanetal01} as described in \citet{Jensenetal03}. This gives an
angular scale of 1\arcs=0.153~kpc. We note that NGC~5898 has redshift
$z$=0.007008 and SBF distance 27~Mpc, but that the uncertainties in the
surface brightness fluctuation measurements are such that the two distances
agree at the 1$\sigma$ level. Uncertainties in our own measurements are
quoted at the 1$\sigma$ level unless otherwise specified in the text.

\section{The NGC~5903 group}
\label{sec:N5903}
\begin{figure}
\includegraphics[width=\columnwidth,viewport=72 115 550 679,clip=]{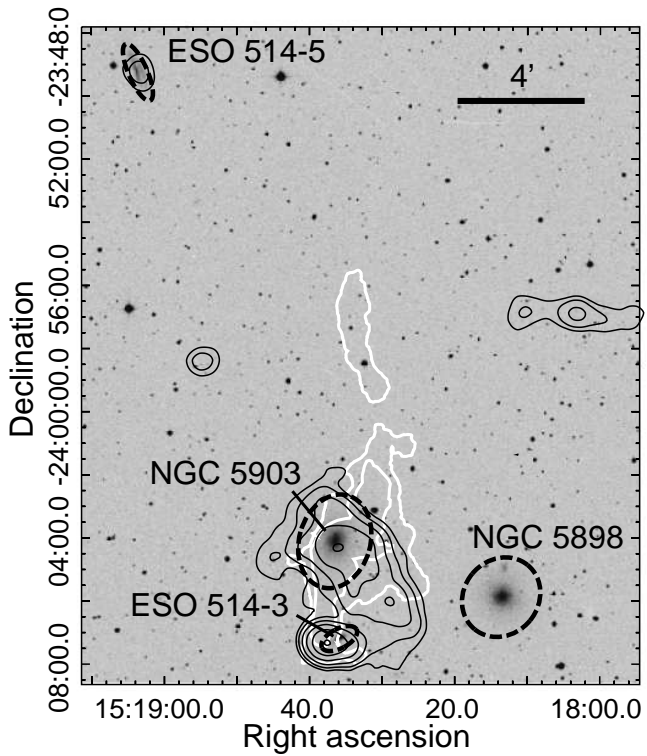}
\caption{\label{fig:wide}Digitized Sky Survey image of the NGC~5903 group. Dashed ellipses mark the \Dtf\ contours of major galaxies. Black contours indicate 1.4~GHz radio emission from the NVSS, starting at 4$\sigma$ significance and increasing in steps of factor 2. White contours show the approximate extent of the \Hi\ structure identified by \protect\citet{Appletonetal90special}, based on their Fig.~1. The two levels roughly correspond to their 0.6 and 1.7$\times$10$^{20}$~atom~cm$^{-2}$ contours.}
\end{figure}

The central part of the NGC~5903 group is shown in Figure~\ref{fig:wide}, with major galaxies marked and 1.4~GHz NRAO VLA Sky Survey \citep[NVSS,][]{Condonetal98} contours overlaid. Table~\ref{tab:gals} provides information on the morphology, luminosity and recession velocity of the group member galaxies discussed in this paper.

\begin{table}
\caption{\label{tab:gals}Recession velocity, $K_s$ band luminosity, and morphological type of the four galaxies discussed in this paper. Recession velocities and morphologies are drawn from NED\protect\footnotemark. Luminosity is calculated as described in \protect\citet{EllisOSullivan06}.}
\begin{center}
\begin{tabular}{lccc}
\hline
Galaxy & v$_{rec}$ & log(L$_K$) & Morphology \\
 & (\kmps) & (\Lsol) & \\
\hline
NGC~5903  & 2565 & 11.10 & E2 \\
NGC~5898  & 2122 & 11.12 & E0 \\
ESO~514-3 & 2345 & 10.27 & E \\
ESO~514-5 & 2538 & 10.52 & Sa \\
\hline
\end{tabular}
\end{center}

\end{table}

\footnotetext{The NASA/IPAC Extragalactic Database (NED) is operated by the Jet Propulsion Laboratory, California Institute of Technology, under contract with the National Aeronautics and Space Administration.}

NGC~5903 was first identified as hosting a bright radio source using the Owens Valley 90-foot telescopes \cite[][90$\pm$30~mJy at 2.6~GHz]{RogstadEkers69}. Later observations with the Parkes \citep{DisneyWall77} and Ooty \citep{GopalKrishna78} radio telescopes showed that there were two components to the emission, a luminous, steep spectrum, extended ($\sim$6\arcm) component associated with NGC~5903 itself, and a more compact ($\sim$1\arcm) source associated with its smaller elliptical neighbour ESO~514-3. 

Imaging with the VLA \citep[][hereafter APW90]{Appletonetal90special} yielded more details of the nature of the two components, a radio double located in or near ESO~514-3, and an intriguing diffuse source whose brightest point coincides with the core of NGC~5903. At 1.4~GHz (see Figure~\ref{fig:wide}) this diffuse source covers the galaxy and extends $\sim$4\arcm\ to the southwest, leading Appleton et al. to suggest a head-tail morphology. However, higher resolution imaging at 150~MHz based on data from the TIFR GMRT Sky Survey (TGSS) suggests that the region detected by Appleton et al. may only be the brightest part of the source \citep[][hereafter GK12]{GopalKrishnaetal12special}. At lower surface brightnesses east-west extent is up to 8\arcm\ ($\sim$75~kpc), and although NGC~5903 is offset from the centre, the source does not have a clear head-tail or jet-lobe morphology.

APW90 also examined the \Hi\ content of the group, finding a filamentary structure extending $\sim$12\arcm\ ($\sim$110~kpc) north-south across NGC~5903. Figure~\ref{fig:wide} shows the approximate extent of this filament, based on Fig.~1 and 2 of APW90. The filament twists to cross the minor axis of the galaxy on a roughly east-west line, and velocity mapping reveals an \Hi\ disc and infalling material, confirming that at least part of the \Hi\ is bound to the galaxy. A spur of \Hi\ also extends from the northwest side of NGC~5903 toward NGC~5898. Interestingly, the diffuse radio structure is anti-correlated with the \Hi\ filament. Although it extends across the \Hi\ filament in NGC~5903, its brightest parts appear to be enclosed by the ``bay'' formed by the filament and spur. The total mass of \Hi\ in the filaments is $\sim$3$\times$10$^9$\Msol, of which $\sim$30\% is located within NGC~5903.

The origin of the filament appears to be tidal or ram-pressure stripping, with the likely lifespan (based on dynamical arguments) $\sim$200-300~Myr. ESO~514-3, though located close to the southern tip of the filament, has no detectable \Hi\ or H$\textsc{ii}$ content, and is unlikely to be the source of the gas. APW90 argue that ESO514-5, a disturbed Sa galaxy $\sim$16\arcm\ (150~kpc) north of NGC~5903, is the likely \Hi\ donor. This galaxy contains $\sim$5$\times$10$^8$\Msol\ of \Hi\ \citep{Theureauetal98}, making it \Hi\ deficient for its type. However, the VLA \Hi\ data obtained by APW90 did not extend far enough to cover ESO~514-5, leaving its role unconfirmed.

Both of the large ellipticals in the group show signs of past disturbance. Deep optical imaging reveals multiple tidal tails or loops in the outskirts of NGC~5898 \citep{Taletal09}, dust features and a double nucleus \citep{RamosAlmeidaetal12}, and the galaxy also hosts a kinematically decoupled core \citep{BertolaBettoni88}. Both NGC~5898 and NGC~5903 show extended regions of enhanced H$\alpha$+[N\textsc{ii}] emission in their cores \citep{Macchettoetal96}. Stellar population modelling suggests that NGC~5903 contains at least two populations, aged 13 and 5~Gyr, and possibly a third $\sim$1~Gyr old population in the galaxy core \citep{Rickesetal08}. Mid-infrared observations suggest that NGC~5898 also underwent an episode of star formation in the recent past \citep{Panuzzoetal11}, though stellar population modelling finds a luminosity weighted stellar age of $\sim$8~Gyr \citep{Annibalietal07}. Neither of the two large ellipticals, nor ESO~514-3, are classed as optical AGN.

\citet{LongVanspeybroeck83} reported an X-ray detection of NGC~5898 with the \textit{Einstein} IPC, but neither of the two ellipticals, nor the group as a whole, were detected in the \rosat\ All-Sky Survey \citep{Beuingetal99}.

\section{Observations and Data Reduction}
\label{sec:obs}
\subsection{Radio continuum observations}

We performed three sets of observations of the NGC~5903 group with the GMRT. The first of these, project 22\_026, was a dual-band observation with central frequencies 234 and 612~MHz, and $\sim$3~hr time on-source. For both bands, data were collected in 256 channels, with bandwidths of 8 and 32~MHz in the 234 and 612~MHz bands respectively. The second, project 25\_006, consisted of \Hi\ line observations of two fields, one covering the group core and one centred on ESO~514-5, the spiral galaxy to the north. The \Hi\ results of these observations will be discussed in a later paper, but we also used the observations to make maps of the continuum emission centred at 1410~MHz, which we include here. The third, project 30\_020, was a single frequency 608~MHz polarization observation of the group centre, and is described in Section~\ref{sec:pol}. Data reduction and analysis for all radio datasets was carried out using the NRAO Astronomical Image Processing System (\textsc{aips}\footnote{http://www.aips.nrao.edu}). Further details of the datasets are provided in Table~\ref{tab:radio}.

\begin{table*}
\begin{center}
\caption{\label{tab:radio} Summary of the radio observations. The half-power beam width (HBPW) and position angle (PA) listed in column 6 represent the maximum resolution of each dataset. Uniform resolution images, made by convolution with a lower resolution beam, for use in later analysis are described in the text.}
\begin{tabular}{lcccccc}
\hline
Project code & Observation date & Frequency & Bandwidth & Observing time & HPBW, PA & rms noise\\
             &                  & (MHz) & (MHz) & (min) & (\arcs$\times$\arcs, \degree) & ($\mu$Jy/beam) \\
\hline
\multicolumn{7}{l}{\textit{GMRT}}\\
22\_026 & 2012 May 19 & 234  & 16  & 180 & 19.7$\times$11.9, 44.3 & 900 \\
        &             & 612  & 32 & 180 & 12.5$\times$6.7, 51.6 & 205 \\
25\_006 & 2013 Dec 22 & 1410 & 16 & 300 & 3.2$\times$2.4, 178.0 & 50 \\
        & 2015 Feb 20 & 1410 & 16 & 335 & 6.7$\times$3.4, 116.5 & 340 \\
30\_020 & 2016 Apr 30 & 608  & 32 & 320 & 8.0$\times$6.6, 18.18 & 120 \\
\multicolumn{7}{l}{\textit{VLA}}\\
AA0099  & 1989 May 16 & 4860 & 50 & 195 & 3.6$\times$2.5, -65.9 & 17.4 \\
\hline
\end{tabular}
\end{center}
Note: The 2013 Dec 22 observation was centred on ESO~514-5. All other observations were centred on NGC~5903.
\end{table*}

Data reduction for project 22\_026 followed a process similar to that described in \citet{Kolokythasetal15}. The data were edited to remove bad antennas and other obvious defects. The flux density scale set based on the flux calibrators. Bandpass calibration was performed, and channels containing noise spikes were removed. To increase the signal-to-noise ratio per channel while limiting the effects of bandwidth smearing, channels in the 612~MHz (234~MHz) data were averaged to $\sim$2~MHz ($\sim$1.5~MHz). The field of view at each frequency was split into multiple facets for imaging, and after further editing, repeated cycles of deconvolution and self-calibration were performed. The resulting images were then corrected for the primary beam pattern of GMRT. 

We followed a similar data reduction path for project 25\_006. For the
ESO~514-5 observation, one round of phase self-calibration produced a
significantly improved image which was used in the final analysis. Images
were created using the whole dataset, with the Robust weighting parameter
set to zero. Self-calibration of the data from the NGC~5903 field did not
improve the image quality and hence was not used. The diffuse extended
emission was not clear in the initial full-resolution map, and we therefore
made low resolution images.  To improve the image quality, we restricted
the longest uv baseline to 10 k$\lambda$ and included a uvtaper of 20
k$\lambda$ on both axes. The Robust weighting parameter was kept at 0. This
resulted in a map with a beam of 22.33\arcs$\times$16.1\arcs\ with
PA=31.4\degree and rms noise of 0.8~mJy/beam. This map was used in further
analysis by convolving to the common angular resolution of 25\arcs\ used
for the other GMRT radio maps.

The GMRT data from all three projects suffer from contamination by emission from the flat-spectrum BL Lac AP Lib (PKS~1514-24). This is a relatively bright source ($\sim$2~Jy across all frequencies we consider), and its imperfect modelling in the image reconstruction leads to enhanced noise levels, particularly in the 1410~MHz maps made from the project 25\_006 data, where the source falls in the outskirts of the field of view and is poorly characterized.

We checked the positional accuracy of our GMRT images by comparing the positions of point sources in the field of view to known radio sources from the NVSS, and where possible sources identified in other wavebands. In all cases, the accuracy was found to be excellent, with offsets for individual sources typically $\la$2\arcs\ and no systematic shift across the field of view. To test the flux scale of our images we measured the fluxes of known background sources. The resulting spectra were generally well described by power laws with spectral index $\alpha$$\sim$0.6-0.9, confirming the flux scales to be accurate.

In addition to our GMRT pointings, we also reprocessed an archival VLA observation, project code AA0099, with central frequency 4.8~GHz. This was performed in BnC configuration, and provides a high spatial resolution view of the continuum sources in the group core. The properties of the dataset are shown in Table~\ref{tab:radio}. Data reduction followed the standard approach for VLA analysis. 

We also make use of two survey images of the group: the 1.4~GHz NVSS image with resolution (HPBW) $\sim$45\arcs\ and r.m.s. noise level 0.5~mJy/bm, and the 150~MHz TGSS image\footnote{downloaded from http://tgss.ncra.tifr.res.in} with resolution $\sim$20\arcs\ and r.m.s. noise 3.5~mJy/bm. We note that the TGSS Alternative Data Release~1 \citep[TGSS-ADR,][]{Intemaetal17} covers the NGC~5903 group, providing images based on the same 150~MHz data, processed using an independent pipeline. Examination showed that for our field, the noise level of the TGSS-ADR images was comparable to the standard TGSS, and the fluxes measured for our sources agreed within $\sim$1$\sigma$ confidence.

\subsection{GMRT polarization observation}
\label{sec:pol}
To further understand interactions in the NGC 5903 group we observed it in the full polar mode of GMRT at 608~MHz. These observations were performed on 2016 April 30, including runs on three phase calibrators (3C~468.1, PKS~1436-16, B2~1308+32) and the amplitude calibrator (3C~286). We used a bandwidth of 32 MHz with 256 channels.

These data were reduced using the standard procedures in \textsc{aips}; in brief this consisted of excision of corrupted data, gain calibration followed by bandpass calibration. The multi-channel data were then subjected to polarization calibration using the procedure outlined in the GMRT polarization analysis guide\footnote{http://www.gmrt.ncra.tifr.res.in/$\sim$astrosupp/docs/aips-pol-gmrt-v2.pdf}. 3C~286 was used as the polarization calibrator. The band-averaged polarized flux density of 3C~286 at 608 MHz was found to be 573~mJy. For an expected total flux density of 20.7~Jy this corresponds to a polarized fraction of 2.78~per~cent, in good agreement with the \citet{Farnes12} estimate of 2.73~per~cent.. We made polarized and Stokes~I images with varying resolution to check for emission. For a map of angular resolution 8\arcs$\times$6.6\arcs, the image had an rms noise of 120~$\mu$Jy and for a map of angular resolution 24.5\arcs$\times$19.5\arcs, the image had an rms noise of 350~$\mu$Jy. There was a considerable increase in noise when the shorter baselines were included in the imaging, but since we were looking for polarization from an extended source these baselines could not be excluded.  

The polarized flux images of the group and its surroundings revealed few sources. No polarized emission was found in NGC~5903 or in the diffuse radio structure. Based on the sensitivity of our images, we estimate the 610~MHz polarization fraction of the diffuse structure to be $\le$10\% (5$\sigma$ upper limit, for r.m.s. noise 120$\mu$Jy/bm). A point source was found coincident with the western half of the double source near ESO~514-3, with polarized flux 5.2$\pm$0.4~mJy, suggesting a polarization fraction $\sim$4.7\%. The background source AP Lib was also detected. We do not consider these likely to be false detections caused by instrumental errors. Such errors are most common outside the 10 per cent power radius of the Stokes $I$ beam \citep[$\sim$35\arcm]{Farnesetal14} and both sources fall well within this radius (ESO~514-3 is only $\sim$3\arcm\ from the beam centre). However, since they have limited impact on our study of the NGC~5903 group, we will not discuss them further.

\subsection{X-ray observations}
\label{sec:Xobs}
We observed NGC~5903 with both \xmm\ and \chandra. Table~\ref{tab:Xray}
gives a summary of the observations. Throughout our analysis we adopt the
Galactic hydrogen column \nh=8.02$\times$10$^{20}$\pcmsq\ \citep[drawn from
the Leiden-Argentine-Bonn survey,][]{Kalberlaetal05}. Spectral fitting was
carried out in \textsc{xspec} v12.8.0, using the solar abundance ratios of
\citet{GrevesseSauval98}. Image and 1-D fitting was carried out in
\textsc{ciao sherpa} v4.7 \citep{Freemanetal01}.

\begin{table*}
\caption{\label{tab:Xray}Summary of the X-ray observations}
\begin{center}
\begin{tabular}{lcccccc}
\hline
ObsID & Observation date & Observing mode & Filter & \multicolumn{3}{c}{Cleaned exposure (ks)} \\
 & & & & MOS & PN & ACIS \\
\hline
\multicolumn{7}{l}{\textit{XMM-Newton}} \\
0722270101 & 2013 July 30-31 & Full Frame & Medium & 48.7 & 41.4 & - \\
\multicolumn{7}{l}{\textit{Chandra}} \\
17152 & 2015 June 26 & VFAINT & - & - & - & 67.1 \\
17674 & 2015 June 28 & VFAINT & - & - & - & 87.9 \\
\hline
\end{tabular}
\end{center}
\end{table*}

\subsubsection{XMM-Newton}
\label{sec:XMM}
NGC~5903 was observed by \xmm\ on 2013 July 30-31 (ObsID 0722270101) for a
total of 51.6~ks. The EPIC instruments operated in full frame mode with
the medium optical blocking filter. Reduction and analysis was performed
using the \xmms\ Science Analysis System (\textsc{sas} v13.0.1). CCDs~3 and 6 of the EPIC-MOS1 detector were inactive owing to micrometeorite damage, and rows 0-149 of CCD~4 were excluded owing to increased noise levels associated with the impact on CCD~3.

In preparation for an initial imaging analysis, we excluded times when the
count rate in any of three bands (10-15~keV, 2-5~keV and 0.3-1~keV)
deviated from the mean by more than 3$\sigma$. This removed a short period
of high background levels at the end of the observation, leaving useful
exposures of 48.7~ks (EPIC-MOS) and 41.4~ks (EPIC-pn). However, soft--band
lightcurves reveal an enhanced background level and increased background
variation, indicative of low-level background flaring through $\sim$75\% of
the observation. 
The group is also located on a
line of sight that passes through diffuse soft X-ray emission in our own
Galaxy.

Bad pixels and columns were identified and removed, and the events lists
filtered to include only those events with FLAG = 0 and patterns 0-12 (for
the MOS cameras) or 0-4 (pn). Point sources were identified using
\textsc{edetect$\_$chain}, and regions corresponding to the 85 per cent
encircled energy radius of each source were excluded. Bright sources
located in or near the optical centroids of group member galaxies 
were retained.

Since the observation is affected by both Galactic foreground emission and
low-level flaring, we chose to perform spectral extraction and fitting using the \xmms\
Extended Source Analysis Software (ESAS) and the background spectral model suggested
in \citet{Snowdenetal04}. The background model consists of: 1) a powerlaw
representing the particle component of the background, which is convolved
with the instrument Response Matrix Function (RMF) and a diagonal Auxiliary
Response Function (ARF), since the particles are largely unaffected by the
telescope mirrors and other optical factors; 2) an absorbed powerlaw
representing the cosmic hard X-ray background, whose index is fixed at
$\Gamma$=1.46; 3) one unabsorbed and two absorbed APEC thermal plasma
models with temperatures of 0.1, 0.1 and 0.25 keV, representing emission
from the galaxy, local hot bubble and/or heliosphere; 4) Two Gaussian
components representing the Al K$\alpha$ and Si K$\alpha$ fluorescent
emission lines at 1.49 and 1.75~keV. Absorption was fixed at the Galactic
value, and the parameters of the cosmic,
Galactic and local X-ray background components were tied across the EPIC
cameras, with relative normalization constants varying to represent the
differences in effective area and field of view between the instruments.
The normalization of the particle and fluorescent background components are
allowed to vary independently in each camera, and after initial fits the
energy and width of the fluorescent lines, the index of the particle
component powerlaw, the temperature of the 0.25~keV thermal component, and the normalization of the cosmic hard X-ray background powerlaw
are also allowed to vary. The temperature of the 0.25~keV component rose to $\sim$0.3~keV in every fit we performed, indicating enhanced Galactic foreground emission.

Particle--only spectra for each extraction region on MOS2 and the pn were
also produced, and subtracted during model fitting. MOS1 was excluded, as
the MOS\_BACK task was unable to correctly scale the particle spectra for
the camera, owing to enhanced gold fluorescence features in the spectrum
which could not be corrected for in the absence of CCDs 3 and 6. Fitting
was performed in the 0.3-10~keV range for MOS and 0.4-7.2~keV range for pn.
The broad energy range allows for modelling of the particle and soft
X-ray background components, with the pn energy range reduced to avoid
background features which would require additional model components.
Spectra from multiple regions covering the majority of the field of view
were fitted simultaneously to improve the constraints on the background
components. A \rosat\ All--Sky Survey spectrum extracted from a
0.75-1\degree\ radius annulus centred on NGC~5903 was also included in the
fit, to help constrain the spectrally soft components of the background
model. The particle and fluorescent components were not fitted to the
\rosat\ spectrum.

\subsubsection{Chandra}
\label{sec:Chandra}
NGC~5903 was observed by \chandra\ on 2015 June 26 and 28, for a total of
155~ks. A summary of the \chandra\ mission and instrumentation can be found
in \citet{Weisskopfetal02}. The ACIS-S3 CCD was placed at the focus of the
telescope, and operated in VFAINT mode. We reduced the data using \textsc{ciao} v4.7 \citep{Fruscioneetal06} and CALDB 4.5.9 following techniques similar to
those described in \citet{OSullivanetal07} and the \chandra\ analysis
threads\footnote{http://asc.harvard.edu/ciao/threads/index.html}.

The observation was performed in two parts, with OBSID 17152 having an exposure of 67.1~ks and OBSID 17674 87.9~ks. Neither suffered from significant background flaring. Very faint mode cleaning was applied to both parts of the observation. In general, the two parts were combined for imaging analysis, but spectra were extracted from each event file separately.

Point source identification was performed using the \textsc{wavdetect} task, based on a 0.3-7~keV image and exposure map for the two parts of the observation combined. In general, point sources were excluded from spectral and spatial fitting, except when point sources associated with group member galaxies were being examined. 

Spectra were extracted from each dataset using the \textsc{specextract} task. Owing to the enhanced soft Galactic foreground emission, the standard blank-sky datasets do not provide an accurate estimate of the background for this observation. Analysis of spectra from the S1, S2, S4 and I3 CCDs, after removal of point sources, suggested that the foreground emission is well described by a 0.3~keV, solar abundance, absorbed APEC thermal plasma with emission measure $\sim$10 times higher than that in the blank sky data. This agrees well with the values found in the \xmms\ ESAS modelling.

Extracting spectra from the S3 CCD, we found that acceptable model fits could be achieved by either a) adding a 0.3~keV thermal component with free normalization to the model, or b) correcting the blank-sky background spectra using appropriately scaled residual spectra from S1. These were created by subtracting a blank-sky spectrum from the spectrum of the S1 chip (after point source subtraction) and scaling using the exposure maps for the relevant regions of the S1 and S3 CCDs. We adopted the latter approach throughout our \chandra\ analysis, using the former to check our results. Comparison of our results with those from the \xmms\ ESAS modelling shows good agreement.

\section{Radio continuum imaging and spectral index measurements}
\label{sec:radio_ims}

\begin{figure*}
\includegraphics[width=0.33\textwidth,viewport=50 140 555 660]{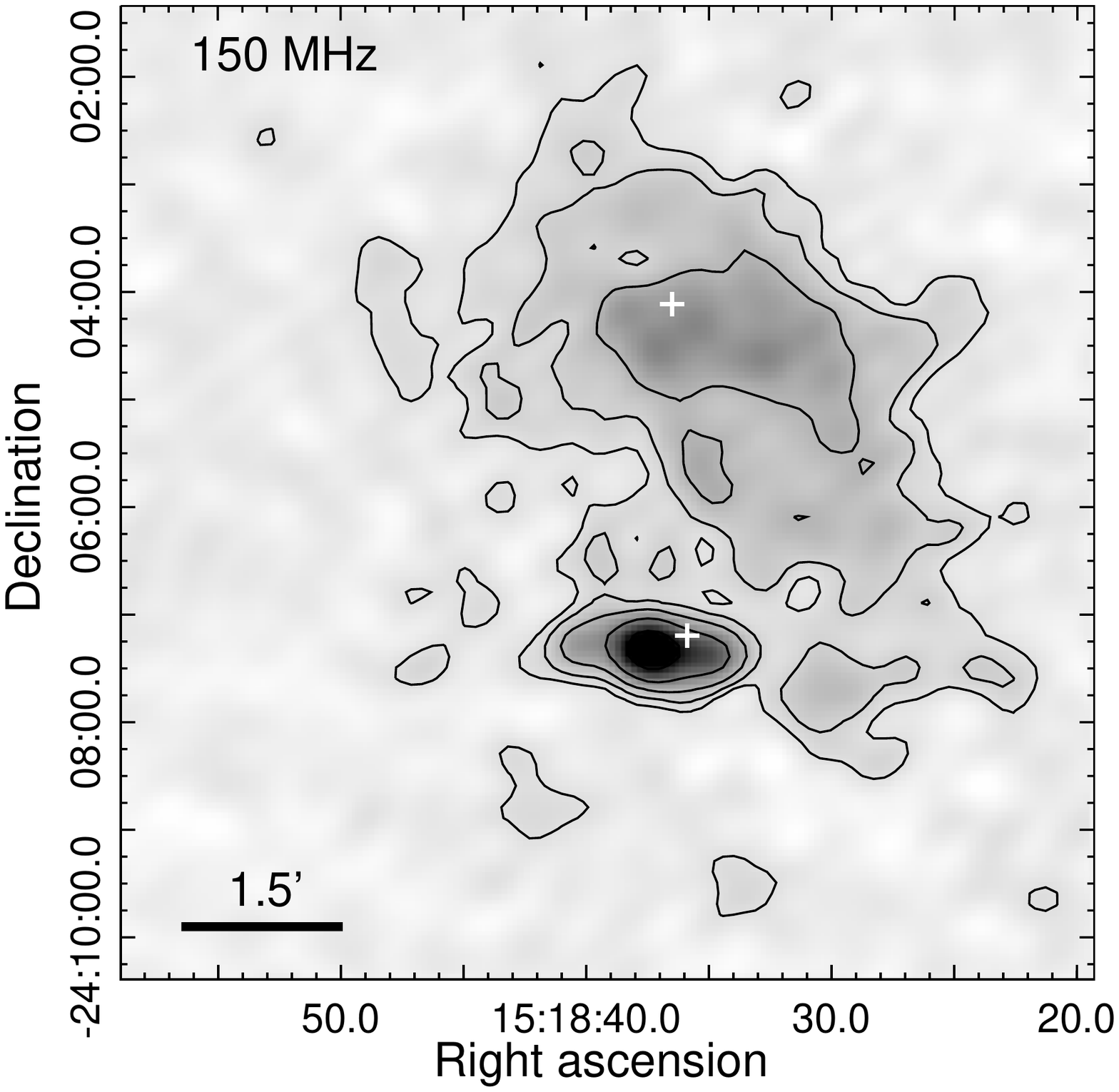}
\includegraphics[width=0.33\textwidth,viewport=50 140 555 660]{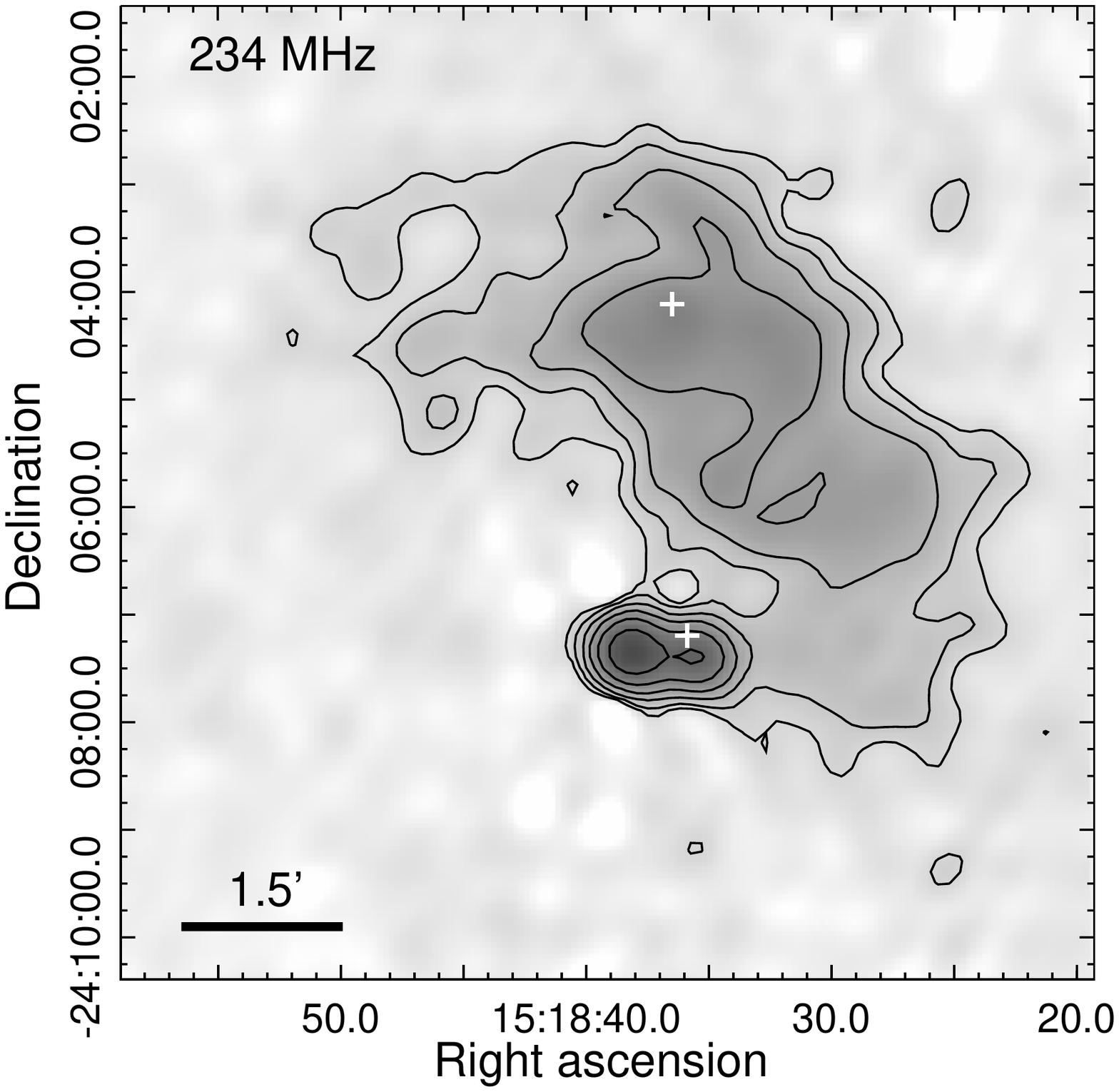}
\includegraphics[width=0.33\textwidth,viewport=50 140 555 660]{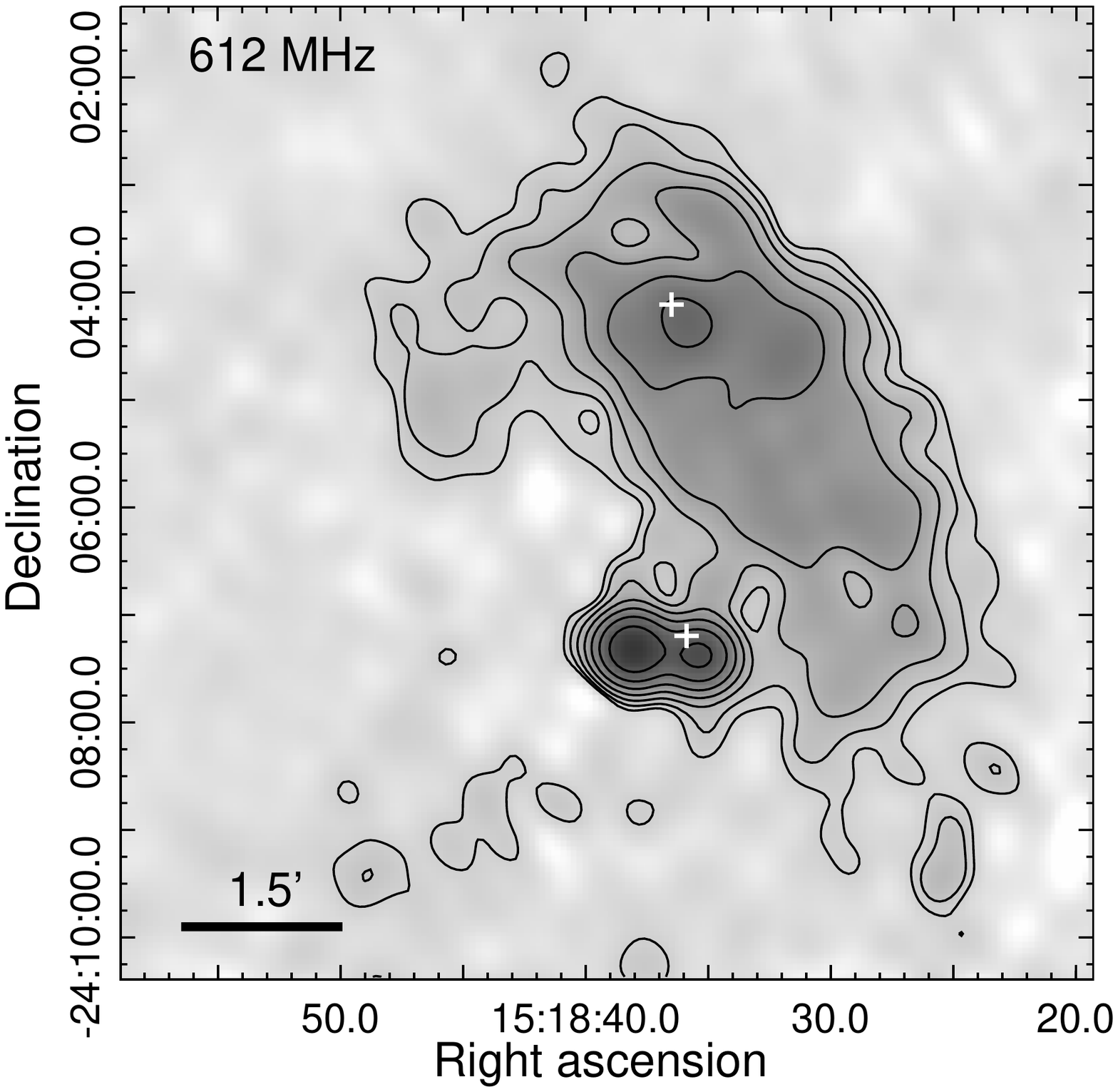}
\includegraphics[width=0.33\textwidth,viewport=50 140 555 660]{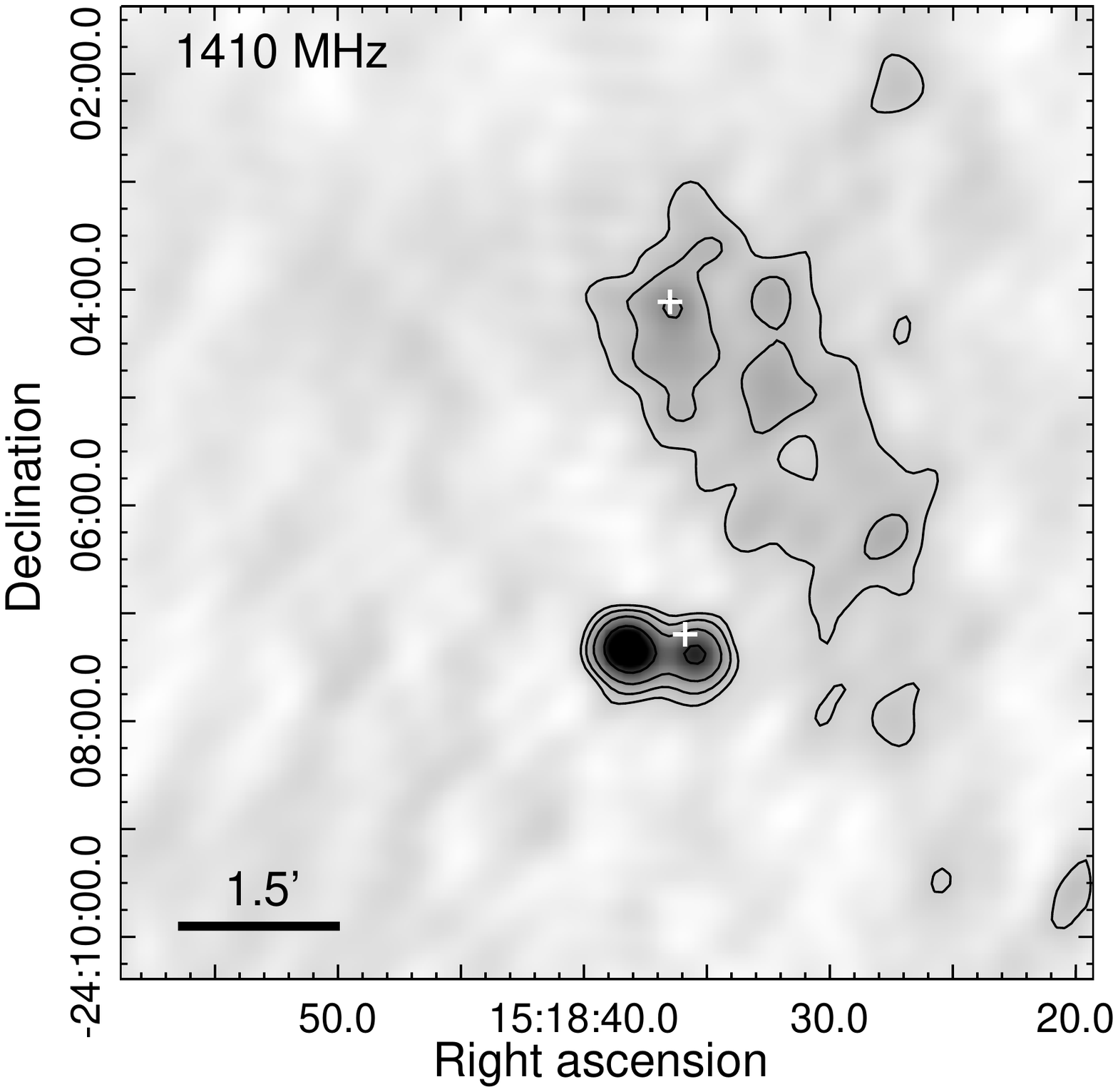}
\includegraphics[width=0.33\textwidth,viewport=50 140 555 660]{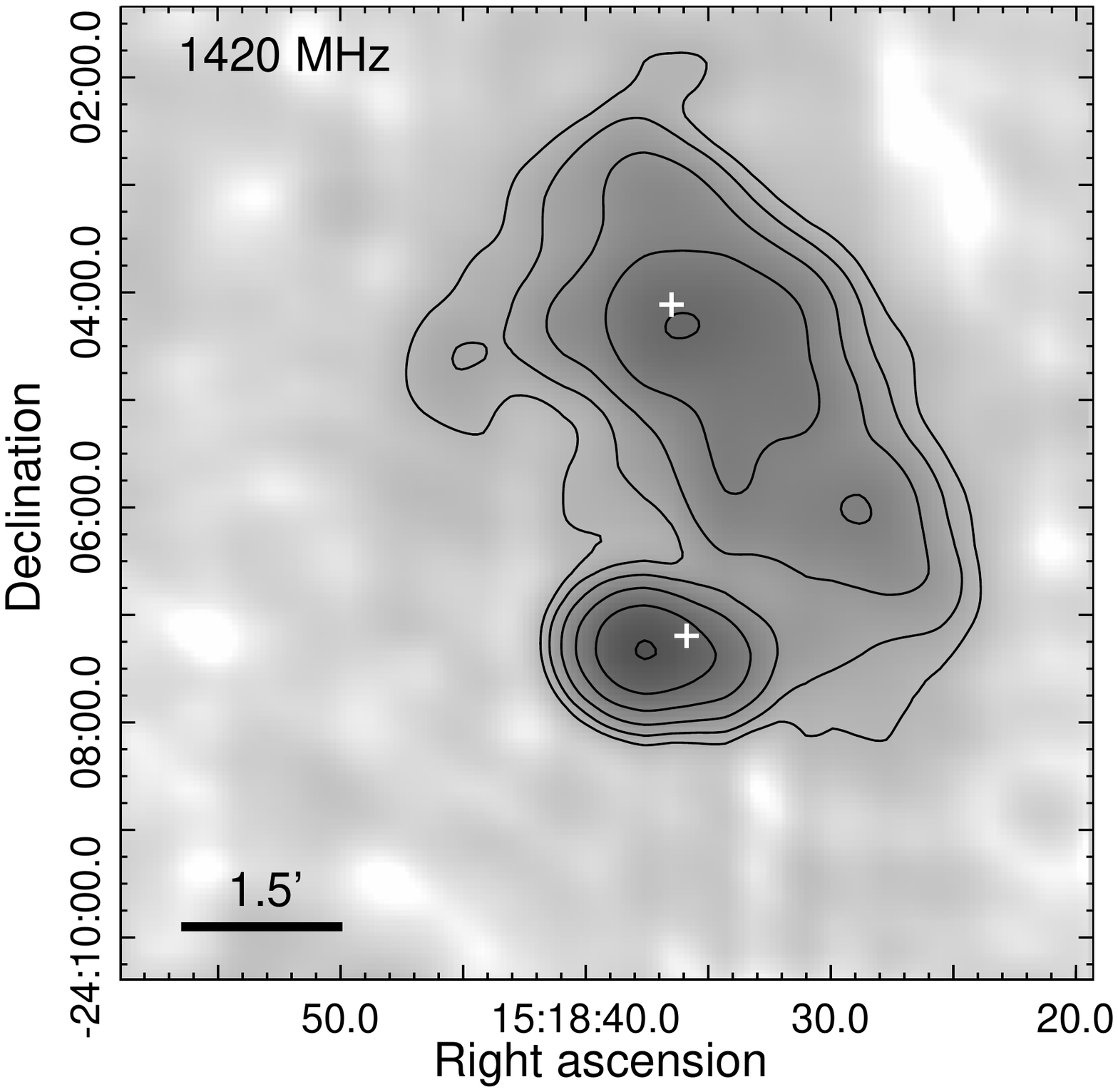}
\caption{\label{fig:GMRT_3panel}Radio continuum images of the NGC~5903 group from the TGSS 150~MHz, our own 234, 612 and 1410~MHz GMRT observations, and the NVSS 1420~MHz survey. The GMRT maps (upper row, and lower left image) are all made with a restoring beam of HPBW 25\arcs. The NVSS image has a restoring beam of HPBW 45\arcs. Contours start at 4$\times$ the r.m.s. noise level (3.5~mJy/bm for the TGSS, 0.5~mJy/bm for the NVSS data) and increase by factors of 2. The centroid positions of NGC~5903 and ESO~514-003 are indicated by crosses.}
\end{figure*}

Figure~\ref{fig:GMRT_3panel} shows TGSS 150~MHz and our GMRT 612 and 234~MHz radio continuum images of the group core. All three images were made using the same circular 25\arcs\ HPBW restoring beam.

The main components of the radio structure are clear at all three frequencies; a bright double source close to ESO~514-3, lying at the southeast edge of a diffuse radio structure $\sim$7\arcm $\times$3\arcm\ across ($\sim$65$\times$27~kpc). The two lobes of the double source are clearest in the 612~MHz band, and the eastern lobe is brighter at every frequency. Neither lobe is centred on the optical centroid of ESO~514-3. 

The brightest part of the diffuse source is located close to NGC~5903, extending to the southwest. The diffuse structure is faintest at its southwestern and northeastern extremities. In the 612 and 234~MHz bands, lower surface-brightness emission extends to the east of NGC~5903. The morphology of the extension differs between the two bands, and at 150~MHz it is less clearly defined, with segments falling below 4$\sigma$ significance. However, the detection across multiple bands confirms that the extension is real.

Figure~\ref{fig:GMRT_3panel} also shows a low-resolution (45\arcs\ HPBW) VLA 1.4~GHz image of the system, and our own GMRT 1.4~GHz image, convolved with a 25\arcs\ HPBW restoring beam. In the NVSS image the two parts of the source near ESO~514-3 are not resolved. The structure of the diffuse emission is very similar to that seen in the lower frequency GMRT images, including an extended arm of low surface-brightness emission east of NGC~5903. Our higher resolution 1.4~GHz image resolves the source near ESO~514-3 more clearly, but only detects the brightest parts of the diffuse structure.

\begin{figure}
\includegraphics[width=\columnwidth,viewport=36 225 576 587]{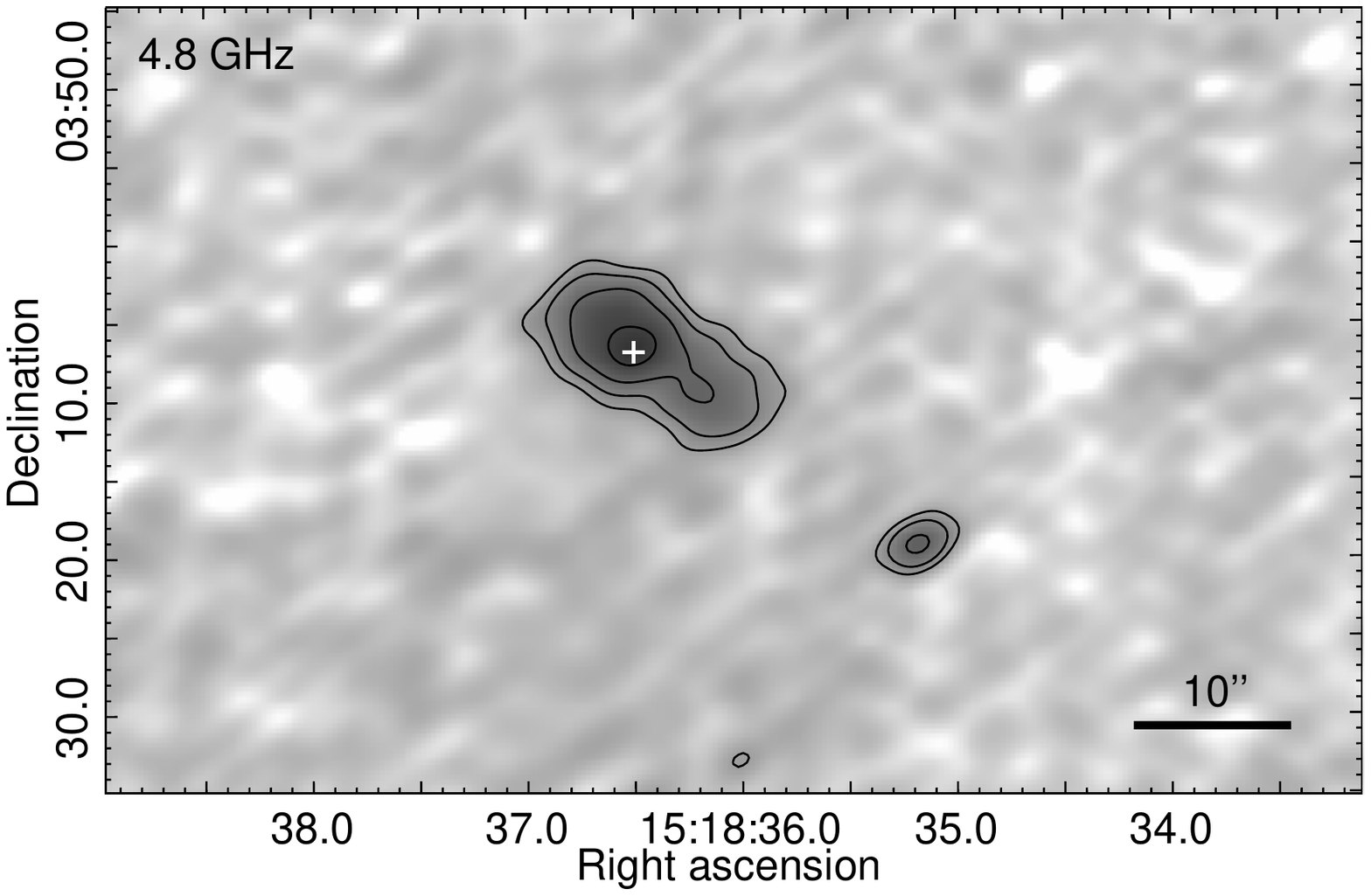}
\includegraphics[width=\columnwidth,viewport=36 228 576 583]{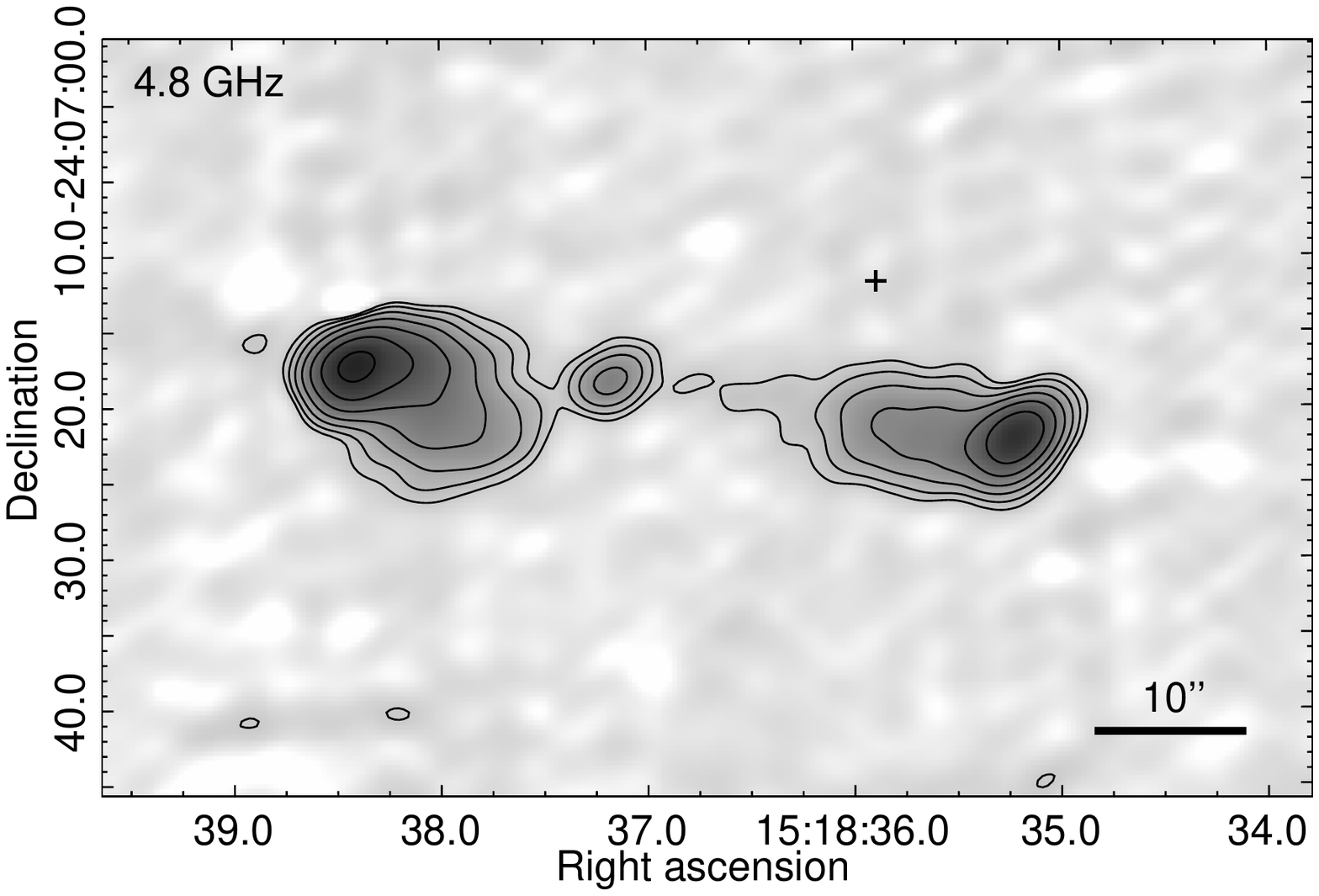}
\caption{\label{fig:VLA5GHz}VLA 4.8~GHz radio continuum images of the sources near NGC~5903 (top) and ESO~514-3 (bottom). Contours begin at 4$\times$ the r.m.s. noise level and increase by factors of 2. The galaxy centroid positions are indicated by crosses.}
\end{figure}

Figure~\ref{fig:VLA5GHz} shows high resolution (3.6$\times$2.5\arcs\ HPBW) VLA 4.8~GHz images of the core of NGC~5903 (upper panel) and the double source near ESO~514-3 (lower panel). The brightest point of the NGC~5903 source coincides with the galaxy optical centroid, and the source is extended along a northeast-southwest axis. This suggests the presence of an AGN with small-scale ($\sim$9\arcs\ or $\sim$1.4~kpc) jets. The angle of the jet axis ($\sim$55\degr\ anti-clockwise from north) is comparable to that of the extended emission ($\sim$35\degr ). The total flux density for core and jets is 4.44$\pm$0.07~mJy. A pointlike source is located to the southwest, along the axis of the central source, and may either be a background object or a clump of radio emitting plasma previously ejected by the AGN. The total flux density of this pointlike source is 470$\pm$17~$\mu$Jy.

The 4.8~GHz image of the source near ESO~514-3 reveals it to have three components rather than the two seen in the lower resolution images; two extended lobes,with a point-like core between them. As seen at lower frequencies, the east lobe is the brighter of the two. The total flux of the source is $\sim$35~mJy, with the core contributing $\sim$1~mJy and the relative fluxes of the east and west lobes having a ratio $\sim$1.6:1. 

Table~\ref{tab:rflux} lists the flux densities measured from the GMRT and VLA data for the source near ESO~514-3 and the diffuse structure. The diffuse structure is resolved out in the 4.8~GHz data, and at lower frequencies the jet source in NGC~5903 is not clearly separated, and so is included in the diffuse flux. For the GMRT 1410~MHz observation, we measure the flux density from the diffuse emission using a region based on the lower-frequency observations. This includes regions outside the lowest contour shown in Figure~\ref{fig:GMRT_3panel}, which may be affected by systematic errors in the image, but is necessary if the measurement is to be comparable with those made at other frequencies. If we instead measure the flux density within a region similar to that enclosed by the outermost (4$\sigma$) contour, we find a flux density of 163.4$\pm$8.2~mJy.

\begin{table}
\begin{center}
\caption{\label{tab:rflux} Radio flux densities for the diffuse radio structure and ESO~514-3. Our own measurements are shown in the upper part of the table, literature measurements in the lower part.}
\begin{tabular}{lccc}
\hline
Telescope & Frequency & \multicolumn{2}{c}{flux density (mJy)} \\
 & (MHz) & Diffuse & ESO~514-3 \\
\hline
GMRT & 150 & 3000$\pm$465 & 993$\pm$154 \\
     & 234 & 1610$\pm$129 & 550$\pm$44 \\
     & 612 & 650$\pm$33 & 256$\pm$13 \\
     & 1410 & 321.5$\pm$16.0 & 101.2$\pm$3.4\\
VLA  & 4860 & - & 34.9$\pm$0.4\\
\multicolumn{4}{l}{\textit{literature measurements}}\\
GMRT & 150$^a$ & 7067$\pm$1100 & 1013$\pm$160 \\
Ooty & 327$^b$ & 1300$\pm$250  & 700$\pm$200 \\
VLA  & 1400$^c$ & 261$\pm$15 & 105$\pm$4 \\
Parkes & 5000$^d$ & 104$^e\pm$30 & - \\
\hline
\end{tabular}
\end{center}
$^a$ \citet{GopalKrishnaetal12special}, $^b$ \citet{GopalKrishna78}, $^c$ NVSS, \citet{Condonetal98}, $^d$ \citet{DisneyWall77}, $^e$ includes flux from ESO~514-3 source.
\end{table}

Table~\ref{tab:rflux} also lists literature flux measurements for the sources. Most interesting is the 150~MHz flux from GK12, which is also drawn from the TGSS data. Gopal-Krishna and collaborators find a much more extended outer envelope to the diffuse structure (see their Fig.~1) with low surface brightness emission extending as far west as NGC~5898, and covering an area $\sim$4\arcm$\times$2\arcm\ to the east of ESO~514-3 and NGC~5903. We therefore consider that our measurements, and the other literature measurements, may only apply to the brighter parts of the diffuse structure.

At higher frequencies, the Ooty and NVSS measurements report separate fluxes for the diffuse emission and the ESO~514-3 source, but the Parkes 5~GHz 4\arcm\ HPBW pencil beam flux does not distinguish between the two \citep{DisneyWall77}. The NVSS fluxes for the two components differ from those measured from our GMRT 1410~MHz observation. It seems probable that the higher spatial resolution of our GMRT observation allows for better separation of the two sources, but it should be noted that the GMRT data are shallower, and affected by noise features caused by AP Lib.

\begin{figure}
\centerline{\includegraphics[width=\columnwidth,viewport=20 220 570 750]{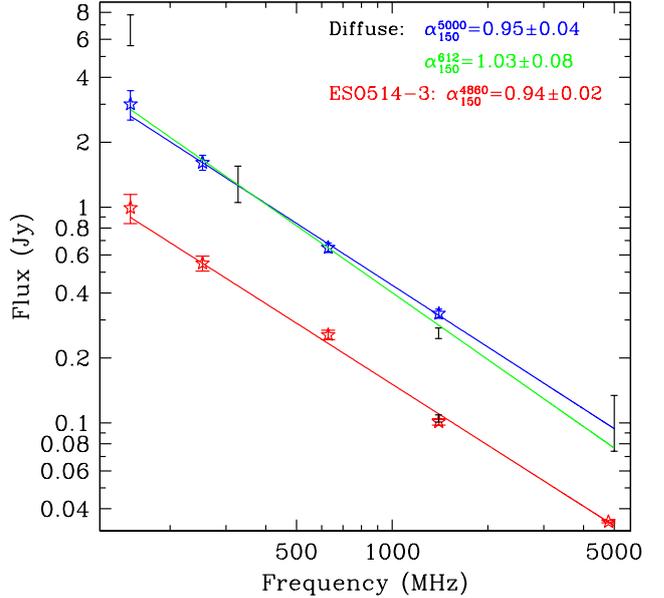}}
\caption{\label{fig:Rspec}Radio continuum spectrum of the NGC~5903 diffuse structure. Our flux measurements are marked by stars with errorbars, literature measurements (see table~\ref{tab:rflux}) by plain errorbars. The three lines represent the fits described in the text, the upper two lines fits to the diffuse component, and the lower line a fit to ESO~514-3.}
\end{figure}

Figure~\ref{fig:Rspec} shows the radio spectra of the diffuse structure and the source near ESO~514-3, using our own measurements and literature data. Our flux estimates for the diffuse structure (or at least its brightest parts) are consistent with a simple powerlaw spectrum. Using our GMRT measurements of the flux at 150, 234, 612 and 1410~MHz with the literature 327 and 5000~MHz measurements, we find a best-fitting spectral index $\alpha_{150}^{5000}$=0.95$\pm$0.04 (where spectral index is defined as $S_\nu \propto \nu^{-\alpha}$). If we exclude the 1.4 and 5~GHz data points, which may suffer from noise or contamination from the ESO~514-3 double source, the best-fitting spectral index is $\alpha_{150}^{612}$=1.03$\pm$0.06. As expected, the 150~MHz flux estimated by GK12, which includes the much more extended low surface brightness component, is inconsistent with these spectral indices. This implies that the extended low surface brightness component detected by GK12 has a much steeper spectral index, $>$1.5.

For the source near ESO~514-3, we find that a simple powerlaw provides a reasonable representation of the spectrum as a whole, with spectral index $\alpha^{4860}_{150}$=0.94$\pm$0.02. However, the 610~MHz point falls above the line and the 1.4~GHz points fall below it. Excluding either of these two frequencies changes the slope, but not by more than the 1$\sigma$ uncertainty. We attempted a fit with a broken powerlaw, but found that when the break frequency was free to fit, the best-fitting value was $\sim$150~MHz. If the break frequency was fixed at 612~MHz, the slopes of the two power laws were identical within the 1$\sigma$ uncertainties. We therefore conclude that the source is best described by a single power law given the data available.

\begin{figure*}
\includegraphics[width=\columnwidth,viewport=36 76 576 715]{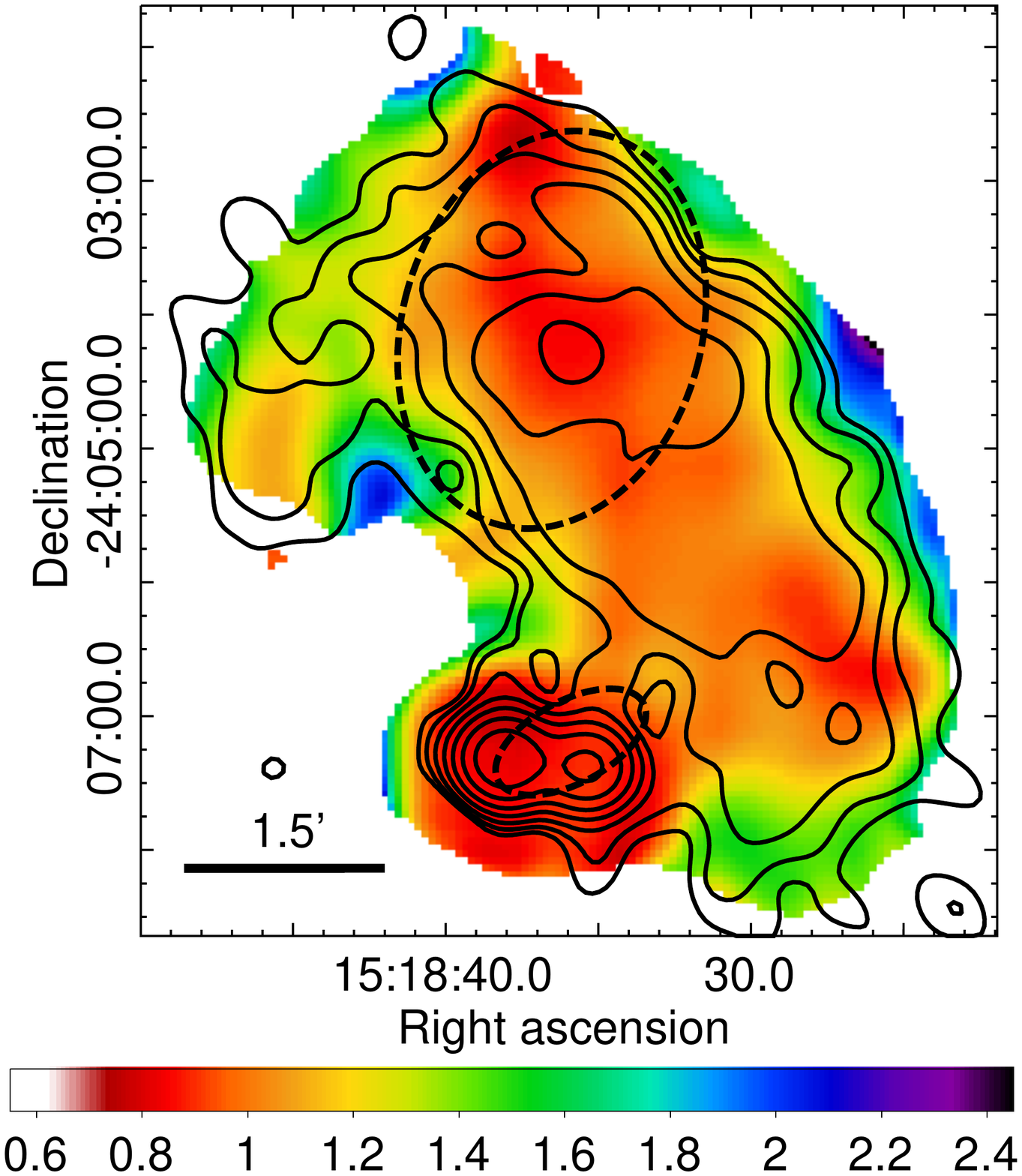} 
\includegraphics[width=\columnwidth,viewport=36 76 576 715]{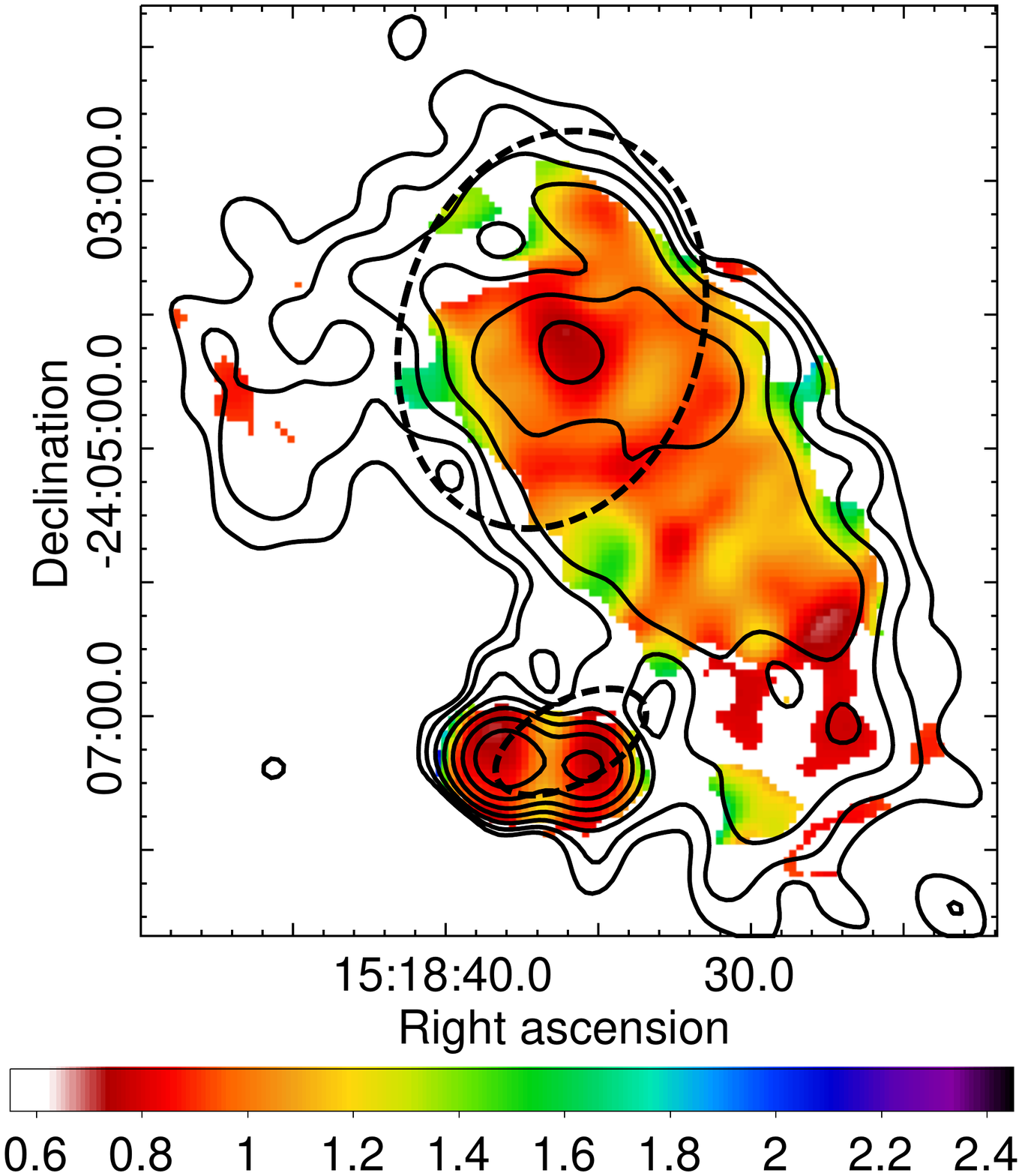}
\caption{\label{fig:multispix}150-1400~MHz spectral index maps, in which the value of each pixel represents the result of a fit to the 150, 234, 612 and 1400~MHz flux density in that region. The left panel uses the NVSS 1.4~GHz image and a 45\arcs\ HPBW restoring beam, while the right panel uses our shallower but higher-resolution GMRT 1.4~GHz image, and a 25\arcs\ beam. Only pixels detected at $>$4$\sigma$ significance in every band, and with spectral index uncertainties $<$5~per~cent are included. GMRT 612~MHz contours based on the 25\arcs\ resolution image are overlaid on both panels, starting at 4$\times$r.m.s. and increasing by factors of 2. The \Dtf\ ellipses of NGC~5903 and ESO~514-3 are also marked. The images have the same angular and colour scales.}
\end{figure*}

In order to examine the spatial variation of the spectral index across the two sources, we created spectral index maps, combining data from the GMRT 150, 234 and 612, and either the VLA 1420~MHz data or the GMRT 1410~MHz observation. The VLA data is deeper but lower-resolution; to match its beam size, images were made in each band using  a 45\arcs\ HPBW restoring beam. The GMRT 1410~MHz data is shallow, but allows the brighter parts of the field to be mapped at 25\arcs\ resolution. In each case we used images with the same pixel scale and orientation, and excluded pixels where the source was detected at $<$4$\sigma$ significance in any of the four bands. We then measured the flux in each band in every remaining pixel. The resulting spectra were fitted with a powerlaw. Figure~\ref{fig:multispix} shows the maps of the resulting spectral indices, filtered to remove all pixels where the uncertainty on the spectral index was $>$5 per cent. 

The flattest indices ($\alpha^{1410}_{150}$=0.73) are found coincident with the nucleus of NGC~5903 and the east lobe of the source near ESO~514-3. Spectral indices $\alpha^{1410}_{150}$$\sim$0.9-1.1 are observed through the brightest parts of the diffuse source, forming a ridge running from NGC~5903 to the southwest. Steeper indices are observed at the edges of the diffuse structure, $\alpha^{1420}_{150}$$\sim$1.2-1.5 in the south, and $\alpha^{1420}_{150}$=1.4-2 on the east and west sides. This is consistent with the picture suggested by the 150~MHz map of GK12, with an extended, steep spectrum, low surface brightness component extending out beyond the regions we detect at 234, 612 and 1420~MHz. The agreement between the two maps is not perfect, as is to be expected given the different resolutions and the relatively shallow 1410~MHz GMRT data. However, they agree well on the general structure of the diffuse source.

When making the maps, we used the full $u-v$ range of each dataset. We note that the extent of the detected emission is smaller than the maximum resolvable scale at every frequency, and we therefore do not expect any biases caused by loss of flux on large scales. The GMRT 1410~MHz observation has the smallest maximum scale, roughly 8\arcm, but we only use the smaller regions detected in that dataset, and find the two maps to be in reasonable agreement, confirming that this dataset does not introduce any bias. While the radio spectra in some pixels are not well described by a powerlaw, this appears to be mainly a product of scatter among the data points, rather than indicating the need for a different spectral model. In particular, the spectra generally do not show a decline at high frequency that would be better fitted by a broken powerlaw or exponential cutoff. We therefore conclude that any break in the spectra occurs at frequencies $<$150~MHz.

\subsection{ESO~514-5}

\begin{figure*}
\includegraphics[width=\columnwidth,viewport=80 160 585 650]{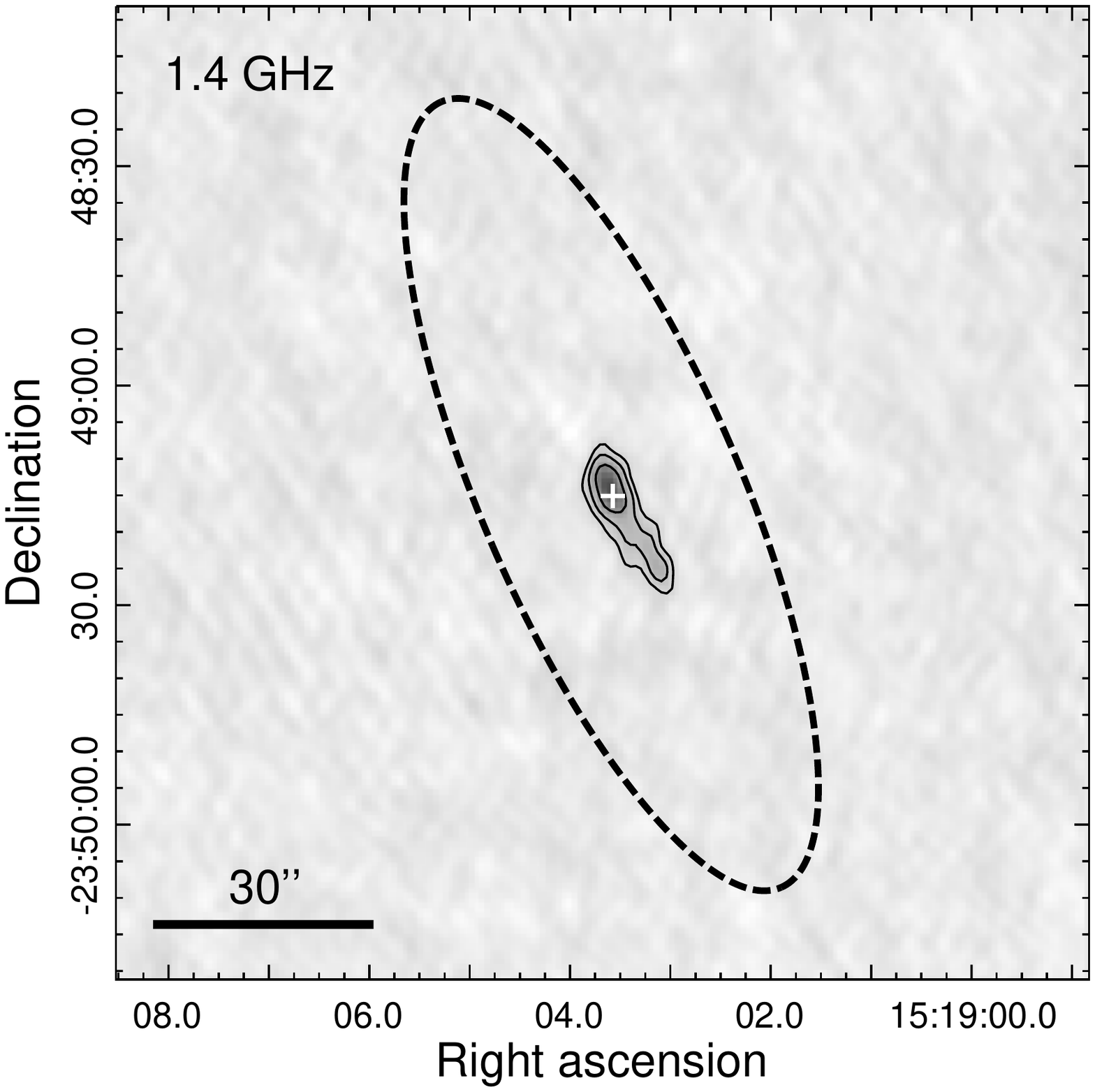}
\includegraphics[width=0.98\columnwidth,viewport=80 150 585 650]{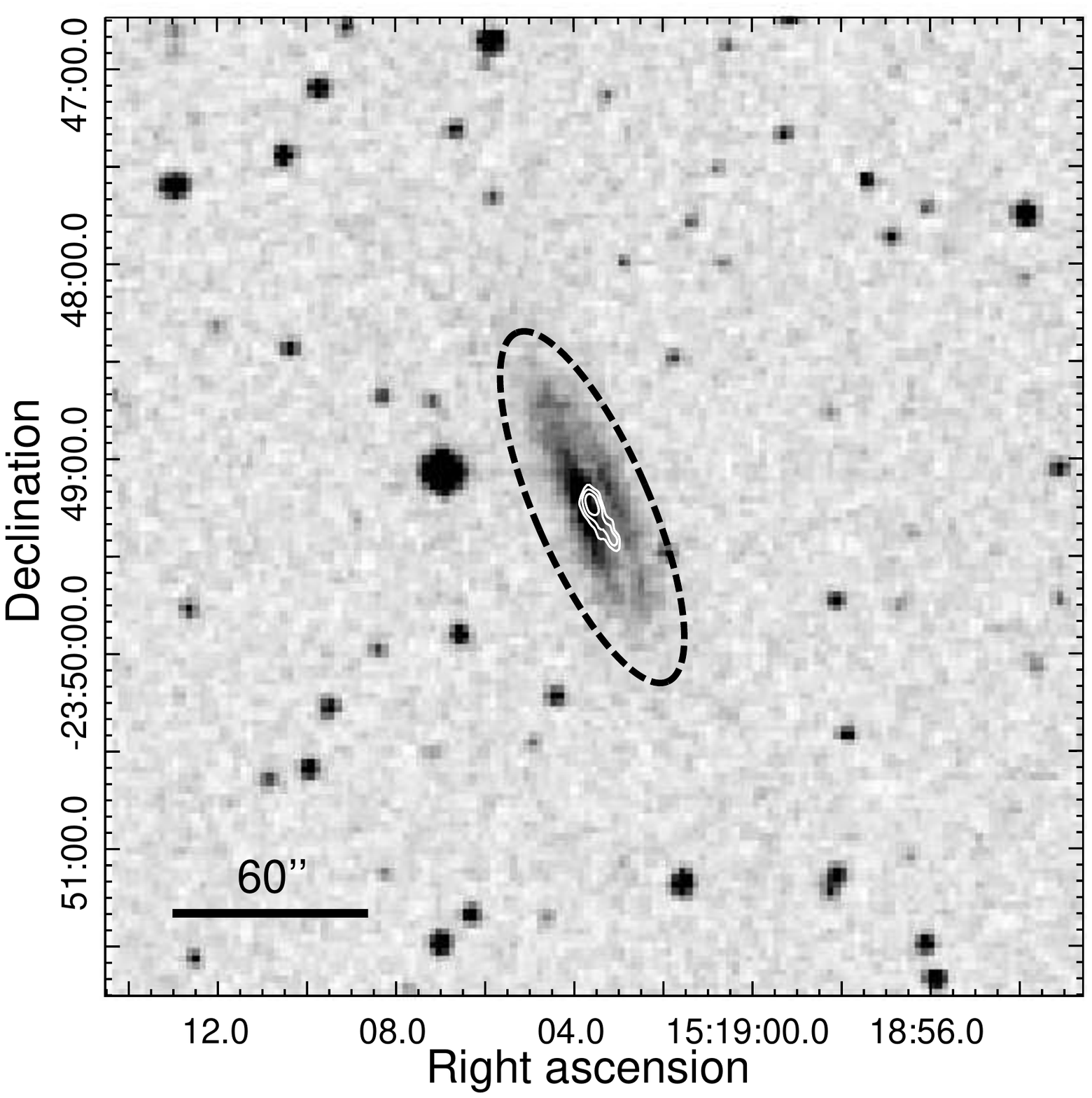}
\caption{\label{fig:e514_1402}GMRT 1.4~GHz full resolution radio continuum image of ESO~514-5 (\textit{left}), and DSS image of the galaxy with 1.4~GHz contours overlaid. Contours begin at 4$\times$ the r.m.s. noise level and increase by factors of 2. The \Dtf\ optical extent of the galaxy is marked by the dashed ellipses, and the galaxy centroid position by a cross.}
\end{figure*}

Figure~\ref{fig:e514_1402} shows the GMRT 1.4~GHz image of the northern spiral galaxy, ESO~514-5, and a DSS image of the galaxy with 1.4~GHz contours overlaid. A point-like source is located at the optical centroid, with emission extending to the southwest along the disc of the galaxy and into the dust lane. The total flux density of the source is 2.64$\pm$0.13~mJy, to which the point source contributes 1$\pm$0.05~mJy. Given the one-sided extension in the plane of the disc, it seems more likely that the emission arises from a combination of an AGN and star formation, rather than a radio jet.

ESO~514-5 is also detected in the lower frequency data, though only as a point source. Table~\ref{tab:E514-5} lists the flux density measurements for the galaxy as a whole at each frequency. The best fitting spectral index of the emission is $\alpha_{234}^{1410}$=0.30$\pm$0.08. One possibility is that the emission is a combination of star formation and AGN emission, dominated by a nucleus with an extremely flat spectrum, as is observed in some other spiral galaxies \citep[e.g.,][]{Mishraetal15}. Higher resolution observations would be needed to confirm this hypothesis.

\begin{table}
\caption{\label{tab:E514-5}Flux density measurements for ESO~514-5, from GMRT observations.}
\begin{center}
\begin{tabular}{lc}
\hline
Frequency & Flux density \\
(MHz) & (mJy) \\
\hline
234  & 5.30$\pm$0.90 \\
612  & 3.08$\pm$0.21 \\
1410 & 2.64$\pm$0.13 \\
\hline
\end{tabular}
\end{center}
\end{table}

\section{X--ray analysis}
\label{sec:xray}
\begin{figure*}
\includegraphics[width=0.495\textwidth,viewport=40 175 570 615]{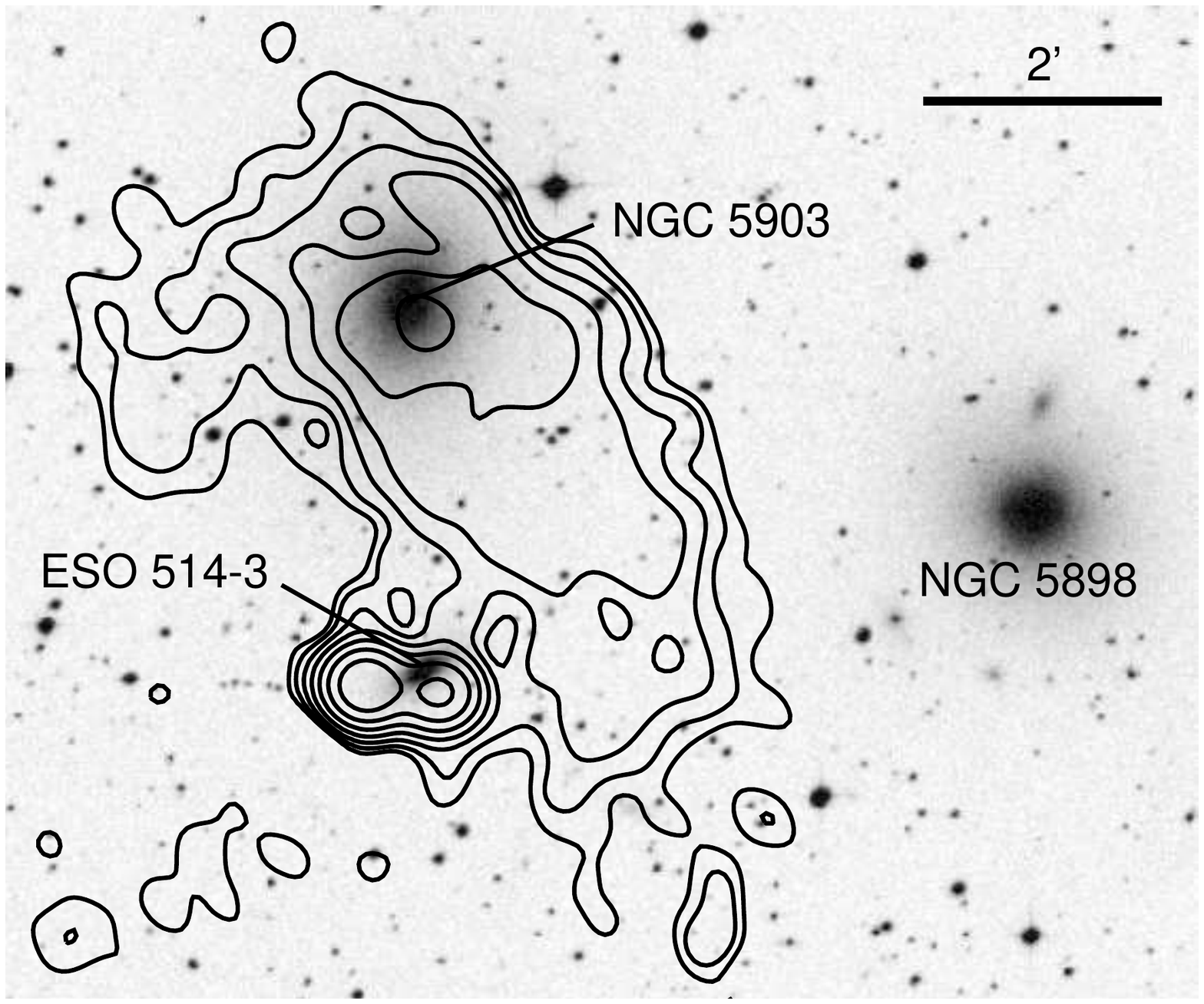}
\includegraphics[width=0.495\textwidth,viewport=40 175 570 615]{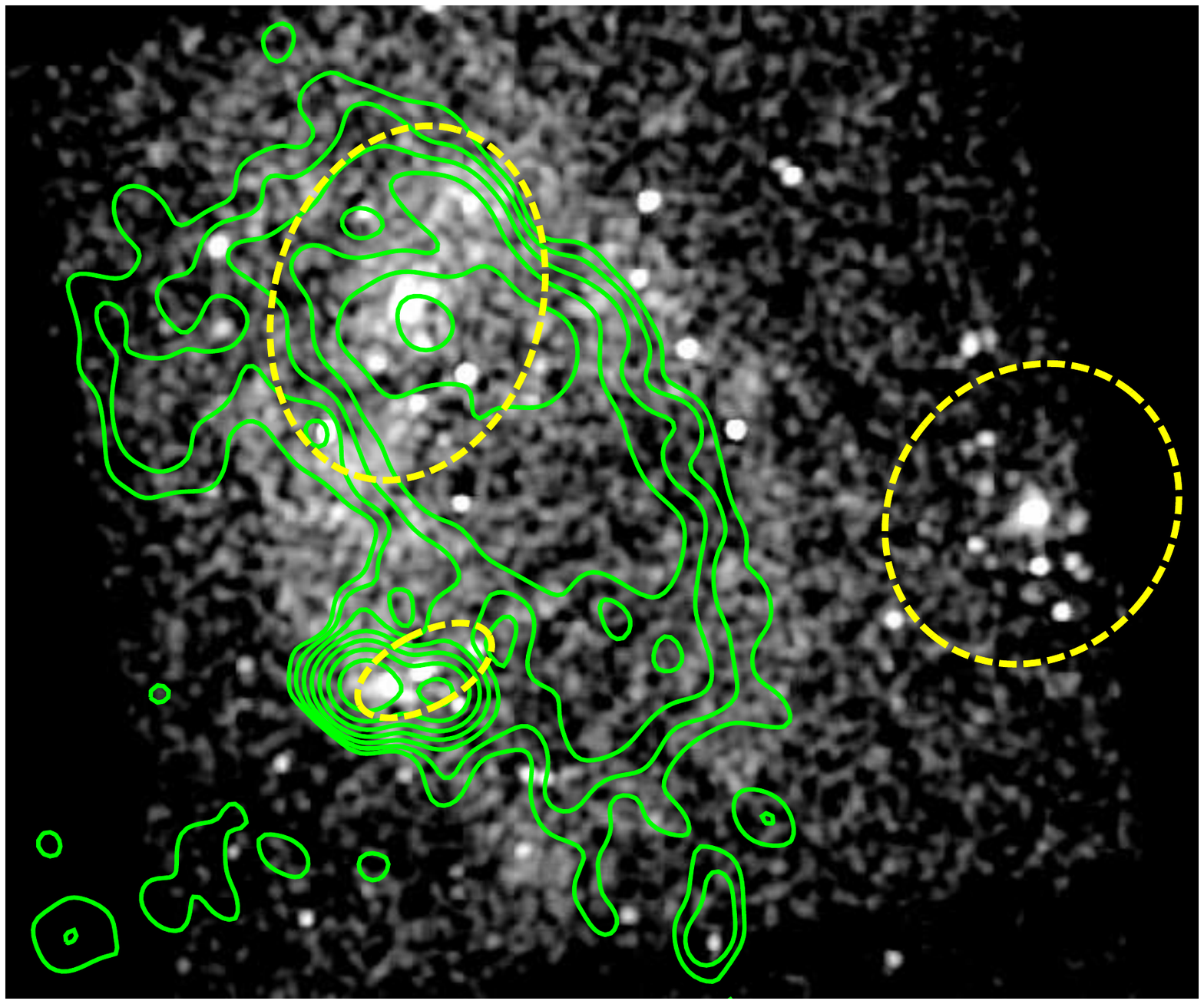}
\includegraphics[width=0.495\textwidth,viewport=40 175 570 615]{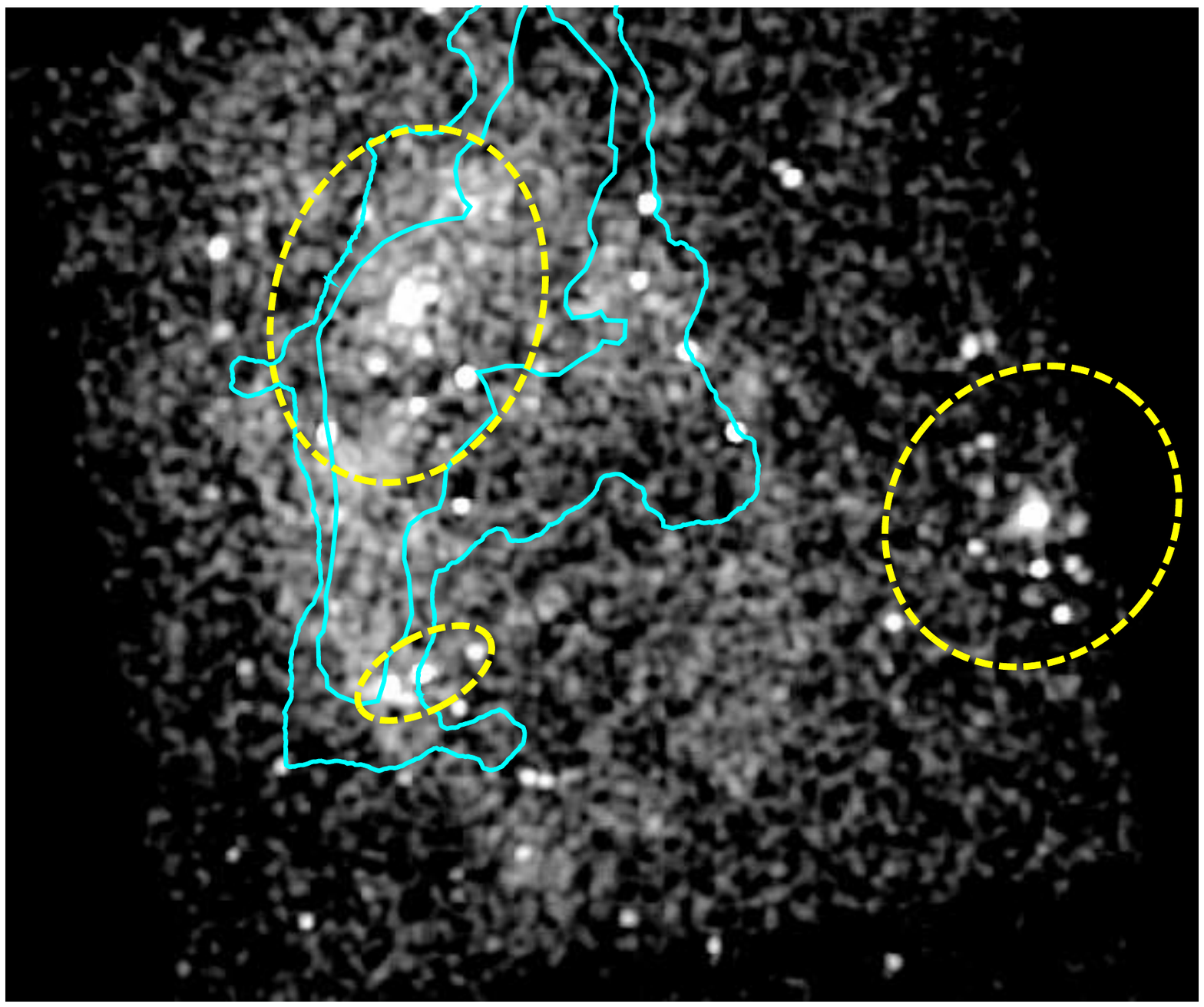}
\includegraphics[width=0.495\textwidth,viewport=40 175 570 615]{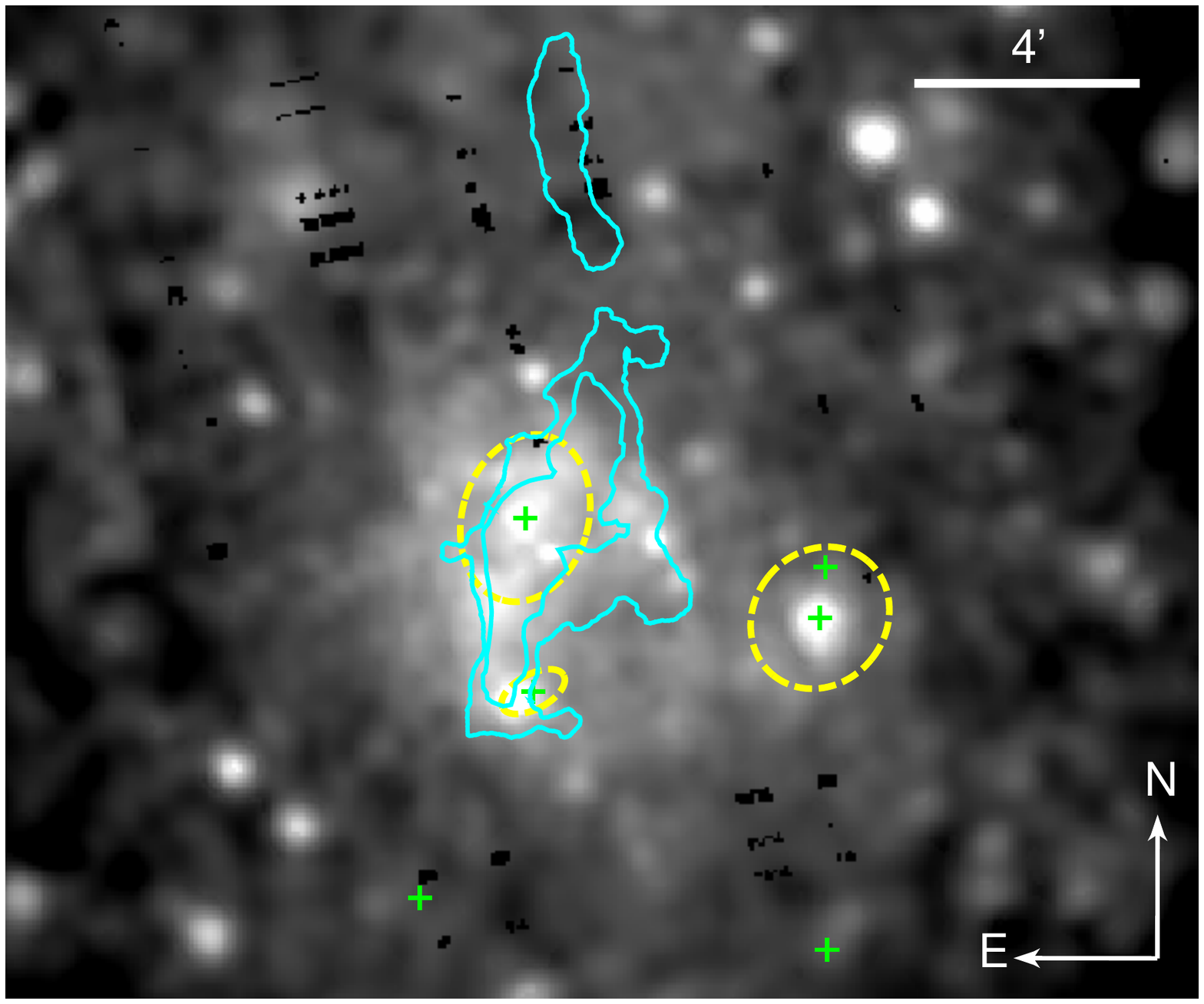}
\caption{\label{fig:Xims}Images of the core of the NGC~5903 group from the Digitized Sky Survey (DSS, \textit{upper left}), \chandra\ (0.5-2~keV, Gaussian smoothed, \textit{upper right} and \textit{lower left}) and \xmms\ (0.3-3~keV, adaptively smoothed, \textit{lower right}). The two upper panels and the lower left panel have the same alignment and scale. In the upper panels, 612~MHz continuum contours, starting at 4$\times$r.m.s. and increasing in steps of factor 2, show the relation between the radio and X-ray emission. In the lower panels, cyan contours show the approximate extent of the \Hi\ emission based on Fig.~1 of APW90, with the two levels roughly corresponding to their 0.6 and 1.7$\times$10$^{20}$~atom~cm$^{-2}$ contours. Dashed ellipses show the approximate \Dtf\ contours of the three major galaxies in the group core, and crosses show the centroid positions of all group members in the field of view.}
\end{figure*}

Figure~\ref{fig:Xims} shows \chandra\ ACIS-S3 and \xmms\ EPIC-MOS+pn
exposure corrected images of the group, with an optical image and 612~MHz
radio contours for comparison. NGC~5903 and NGC~5898 both host a number of
point sources within their \Dtf\ ellipses, the brightest source in each
galaxy coincident with the galaxy centre. ESO~514-3 contains 3 point
sources, one of which is centrally located, but the brightest of the three
is offset to the southeast. The central point sources in each galaxy are
likely AGN, and are discussed further in Section~\ref{sec:AGN}.

Diffuse emission is clearly visible in the group core, with the brightest emission associated with NGC~5903. A roughly elliptical loop of bright emission extends southwest from NGC~5903, overlapping ESO~514-3. The brightest segments of the loop are located between NGC~5903 and ESO~514-3, and to the west of NGC~5903. These regions correspond to the southern part of the \Hi\ filament and its western spur, but the fainter parts of the loop extend beyond the limit of detected \Hi. X-ray emission fills the loop and extends beyond it, particularly to the north and east of NGC~5903. Enhanced emission is visible between the west edge of the loop and NGC~5898, and there appears to be a weak linear feature crossing the loop from the approximate position of ESO~514-3 to NGC~5898.

The surface brightness distribution in NGC~5903 and the X-ray loop is not smooth, and narrow regions of reduced surface brightness are visible within the \Dtf\ ellipse of NGC~5903, between the galaxy core and the base of the two arms of the loop. These features are visible in both \chandra\ and \xmms\ images. These ``channels'' do not correspond to any structures in the radio continuum images. While ESO~514-3 lies on the same line of sight as the X-ray loop, it is unclear whether it is embedded in that structure, or merely superimposed on it. 

Comparing the X-ray and 612~MHz radio emission, we find that the brightest diffuse radio emission (i.e., excluding the double radio source near ESO~514-3) extends across NGC~5903 and fills the loop structure. The southern and western edges of the radio emission appear to be relatively well correlated with the inner edges of the X-ray loop. However, we note that the low surface brightness 150~MHz emission reported by GK12 extends beyond the loop in every direction, particularly to the east. 

Comparison with the radio spectral index maps shows that the radio emission inside the X-ray loop has the flattest spectrum, while steeper emission is found coincident with the edges of the X-ray loop and in the confused X-ray structures to the north and east of NGC~5903. There is no clear correspondence between the X-ray and radio structures in this region, though the radio and X-ray plasmas appear to be colocated. Comparing the GK12 150~MHz map to the \xmms\ image, we find some diffuse X-ray emission to the east of the loop in the region of low surface-brightness (presumably steep spectrum) radio emission, but no clear correlation between structures.

\subsection{Large-scale diffuse X-ray emission}
Diffuse emission extends to the edge of the \xmms\ field of view ($\sim$950\arcs\ or $\sim$145~kpc radius) and appears to have a relatively smooth, symmetric distribution outside the group core. To examine the properties of this extended IGM, we extracted spectra from the EPIC MOS2 and pn, and simultaneously fitted them with an absorbed APEC model, using the ESAS background modelling method. 

The typical temperature of the group outside the disturbed core was $\sim$0.65~keV, and the abundance is low, $\sim$0.25~\Zsol. Including regions covering the galaxies and structures in the group core, we find the total 0.3-7~keV gas luminosity within 950\arcs\ is 1.63$^{+0.06}_{-0.05}\times$10$^{41}$\ergps.

\begin{figure}
\includegraphics[width=\columnwidth,viewport=20 230 570 770]{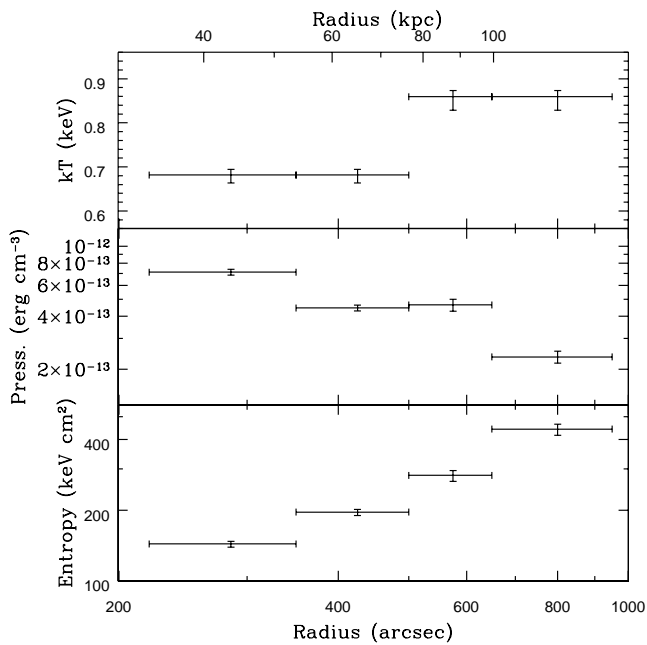}
\vspace{-2mm}
\caption{\label{fig:deproj}Deprojected temperature, pressure and abundance profiles for the NGC~5903 group, outside the disturbed core region.}
\vspace{-5mm}
\end{figure}

Splitting the outer part of the \xmms\ field (220-950\arcs) into four annular bins, we attempted a fit to the diffuse group halo emission with a deprojected absorbed APEC model. Using full 360\degree\ annuli, we found unphysical fluctuation in the pressure profile. We therefore excluded a 70\degree\ sector north of the group core, where the IGM emission appears marginally brighter than in other directions. The low signal-to-noise ratio of the remaining spectra forced us to freeze abundance at 0.25~\Zsol, and tie the temperatures in the inner and outer pairs of bins, but we were able to extract deprojected pressure and entropy profiles. These are shown in Figure~\ref{fig:deproj}. 

The temperature profile shows an increase in temperature at large radius from $\sim$0.7~keV to $\sim$0.9~keV, in conflict with the projected spectral fits. Such an increase is unusual, and raises the possibility that the background model may be oversubtracting soft emission. If the temperature at large radius is overestimated, the true pressure profile will be steeper than shown in Figure~\ref{fig:deproj}, with lower pressures in the outer two bins. This could explain the identical pressures (within errors) in bins 2 and 3. However, we note that if this is the case, the values for the innermost bin should still be reasonably accurate, since the innermost bin has the highest signal-to-noise ratio, and poor fit values in the outer bins will be compensated for in the deprojection model by alterations to the fits in intermediate bins.

\subsection{Temperature distribution in the group core}
To examine the temperature and abundance distribution in the complex core of the group, we divided the \chandra\ S3 field of view into a number of regions based on the structures visible in the optical, X-ray, and radio images. The cores of NGC~5903 and NGC~5898 were approximated with 30\arcs\ radius circles, and ESO~514-3 by its \Dtf\ ellipse ($\sim$37$\times$18.5\arcs). An elliptical annulus of width $\sim$50\arcs, split into four quadrants, was used to approximate the X-ray loop, and the region inside it broken into halves. Partial annular segments in the north side of NGC~5903 were used to examine the bright emission in the northern part of the galaxy, and a rectangular region used to cover any emission between the loop and NGC~5898. The remainder of the field was broken up into large rectangular regions. All point sources were excluded.

\begin{figure*}
\includegraphics[height=0.37\textwidth,viewport=36 92 576 700]{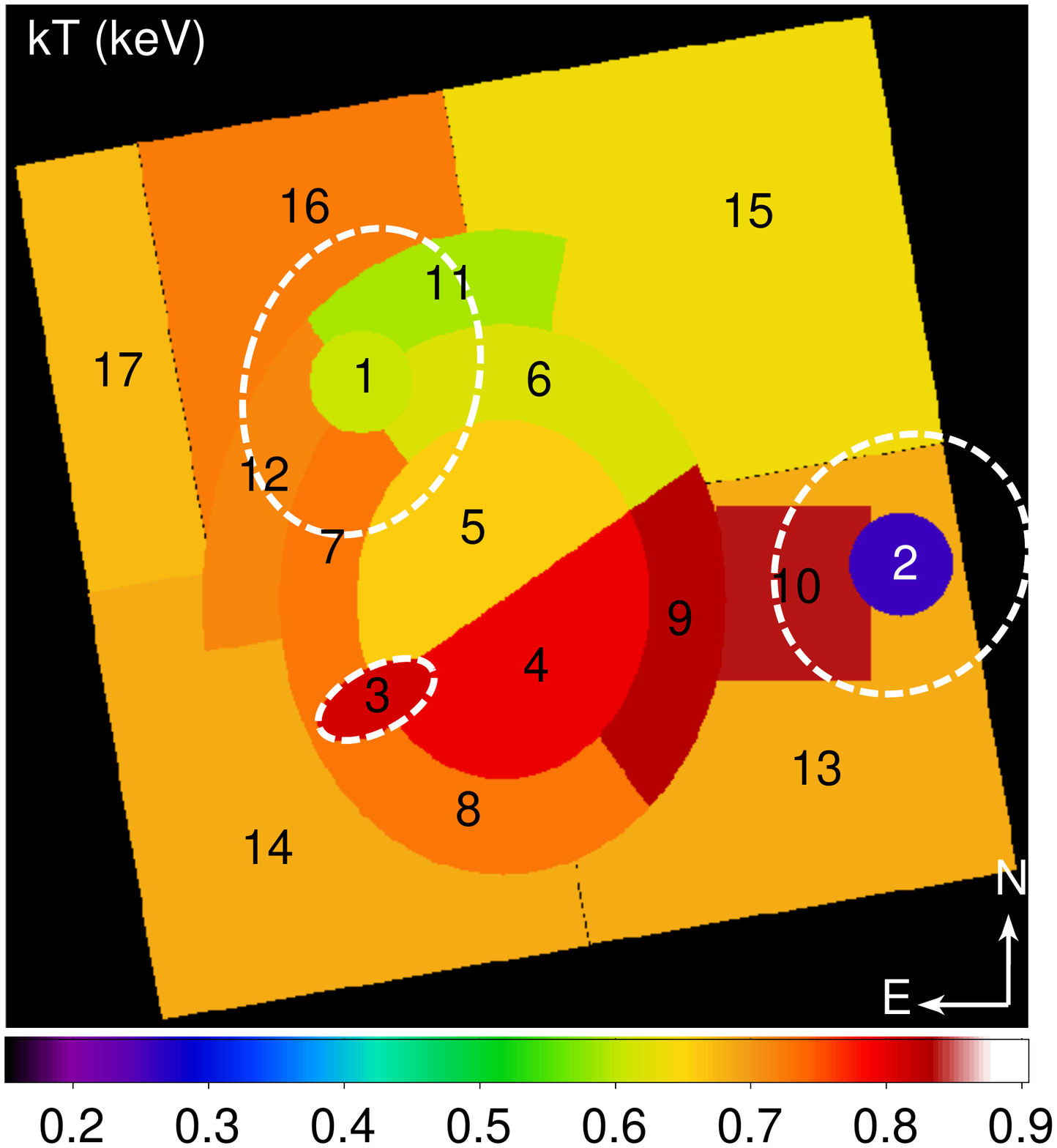}
\includegraphics[height=0.37\textwidth,viewport=36 92 576 700]{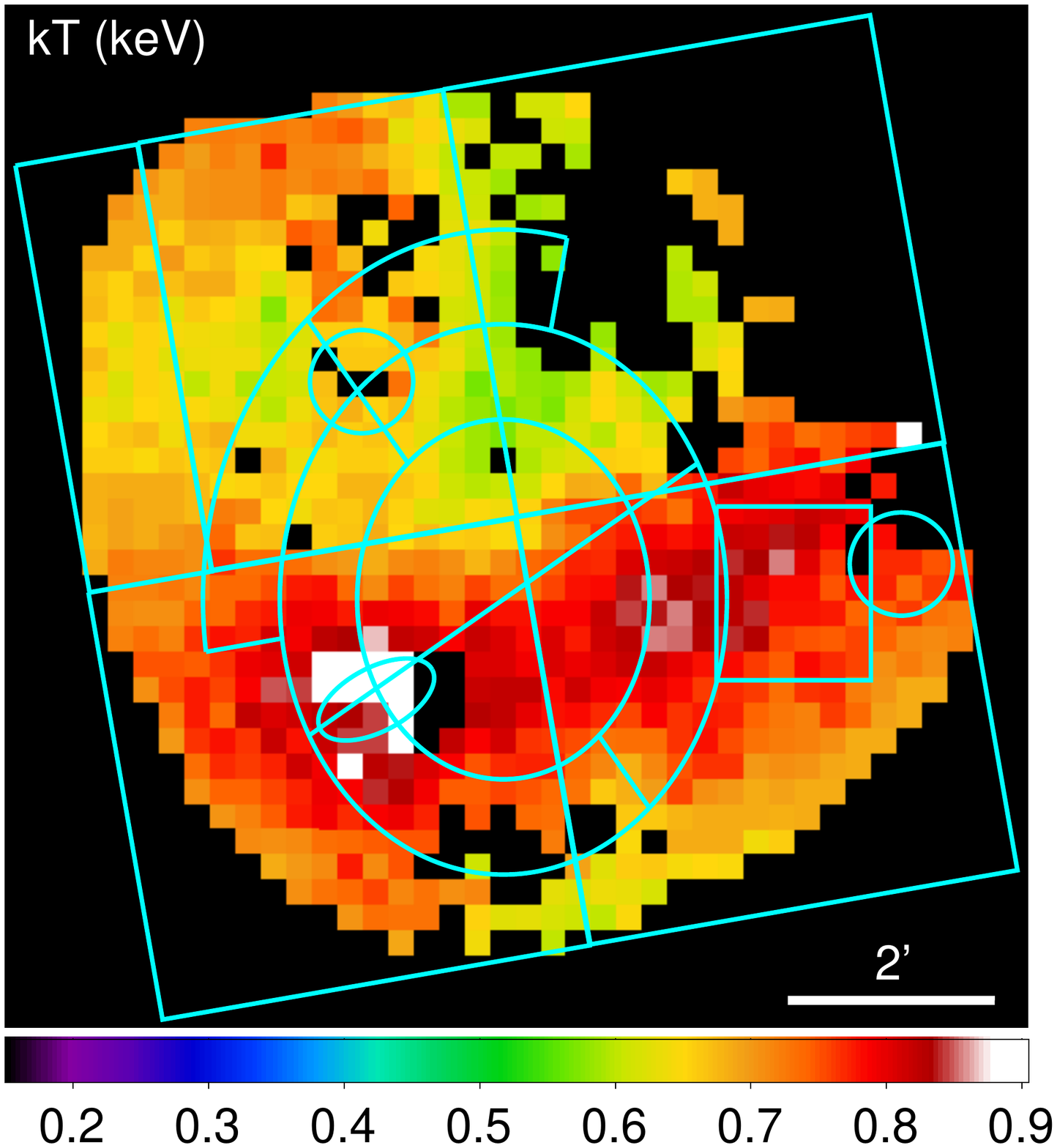}
\includegraphics[height=0.37\textwidth,viewport=36 92 576 700]{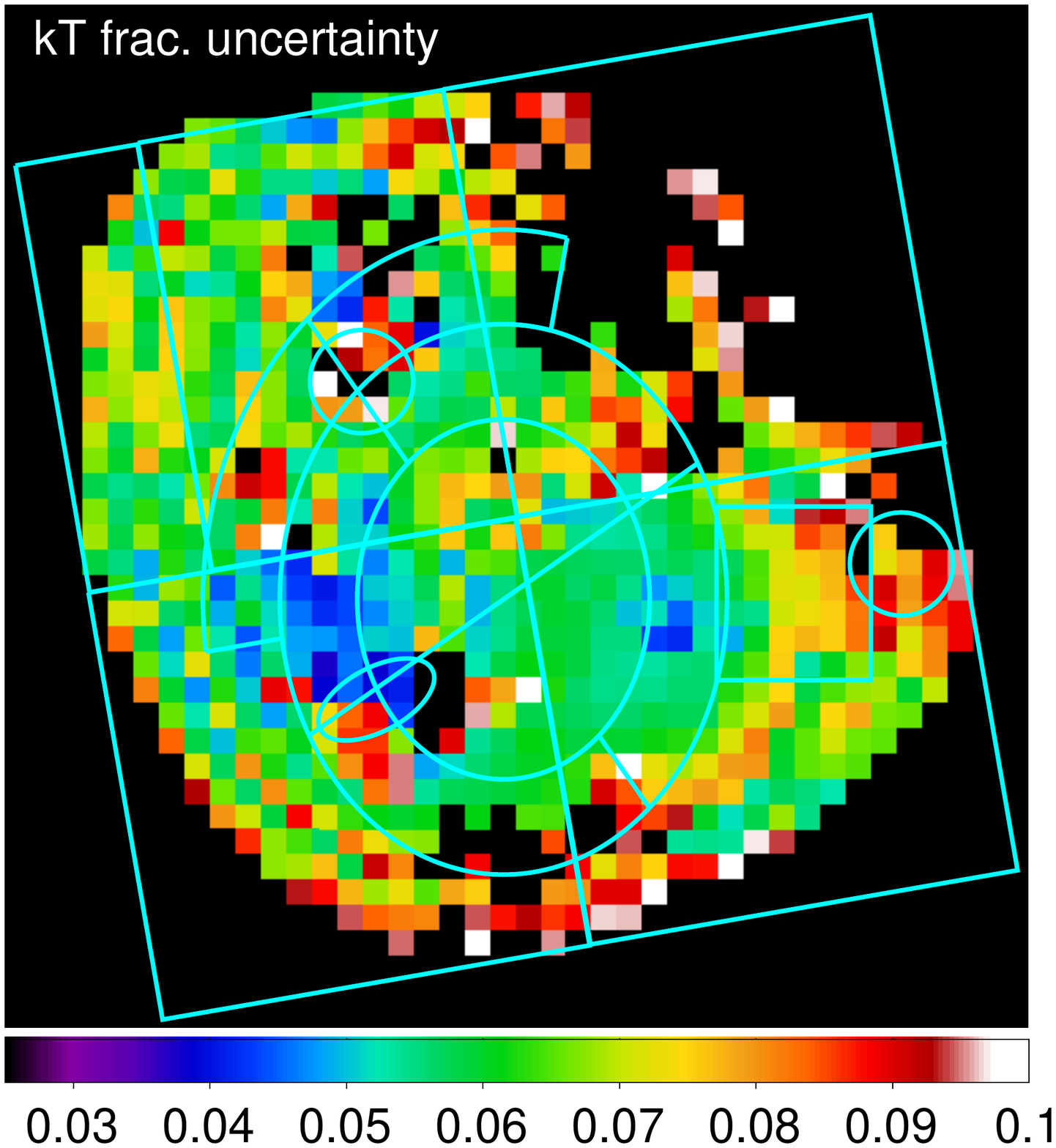}
\caption{\label{fig:Tmap}Maps of best-fitting temperature based on the \chandra\ data. The left panel shows a temperature map using regions selected based on the optical, X-ray and radio structures in the group. The fit parameters and uncertainties for each numbered region can be found in Table~\ref{tab:Tmap}. Galaxy \Dtf\ contours are marked with dashed ellipses, and the map covers the full 8.4\arcm\ square ACIS-S3 field of view. Regions covering galaxy cores were fitted with absorbed APEC+powerlaw models, all other regions with APEC models. The centre panel shows a temperature map generated with fixed 15\arcs\ pixel scale, but allowing spectral extraction regions to vary between 34-94\arcs\ radius. Typical uncertainties are $\sim$6\%, and pixels with uncertainty $>$10\% have been removed. The colour bar is identical to that of the left panel, and region boundaries from that panel are marked in cyan. The right panel shows the fractional uncertainty on the temperature.}
\end{figure*}

Spectra were extracted for each region, and fitted using absorbed APEC or,
where a hard excess above the APEC fit was observed, APEC+powerlaw models.
The APEC+powerlaw fits were required in regions corresponding to the cores
of the galaxies NGC~5903, NGC~5898 and ESO~514-3, so the powerlaw component
likely arises from unresolved point sources in the galaxies. In some
regions abundance was not constrained by the data, and was therefore fixed
at 0.25~\Zsol. The best fitting models for each region are shown in
Table~\ref{tab:Tmap}, and these were used to create the map of temperature
shown in the left panel of Figure~\ref{fig:Tmap}.

\begin{table*}
\caption{\label{tab:Tmap}Fit parameters for each of the regions in the temperature map shown in Figure~\ref{fig:Tmap}. A powerlaw component was included in those regions where a hard excess above the APEC model was observed. Abundance values without uncertainties indicate that the parameter was fixed, since it was not constrained by the data. The final column gives the reduced $\chi^2$ value for the fit, and the degrees of freedom.}
\begin{center}
\begin{tabular}{lcccccc}
\hline
Region & Net counts & Model & kT & Z & $\Gamma$ & red. $\chi^2$ / d.o.f. \\
 & (0.5-5~keV) & & (keV) & (\Zsol) & & \\
\hline
1 & 819 & APEC+PL & 0.61$\pm$0.05 & 0.25 & 1.57$^{+0.20}_{-0.29}$ & 0.78 / 62 \\
2 & 413 & APEC+PL & 0.26$^{+0.10}_{-0.08}$ & 0.25 & 2.62$^{+0.48}_{-0.41}$ & 0.47 / 35 \\
3 & 496 & APEC+PL & 0.81$^{+0.12}_{-0.09}$ & 0.25 & 2.12$\pm$0.44 & 1.28 / 36 \\
4 & 1606 & APEC & 0.66$^{+0.05}_{-0.04}$ & 0.24$^{+0.10}_{-0.07}$ & - & 0.79 / 162 \\
5 & 1129 & APEC & 0.79$\pm$0.06 & 0.20$^{+0.12}_{-0.08}$ & - & 0.85 / 145 \\
6 & 1569 & APEC & 0.62$\pm$0.03 & 0.23$^{+0.10}_{-0.06}$ & - & 0.64 / 137 \\
7 & 1479 & APEC & 0.73$^{+0.03}_{-0.04}$ & 0.31$^{+0.13}_{-0.08}$ & - & 0.75 / 110 \\
8 & 932 & APEC & 0.73$^{+0.05}_{-0.06}$ & 0.23$^{+0.15}_{-0.09}$ & - & 0.69 / 111 \\
9 & 879 &  APEC & 0.83$\pm$0.05 & 0.47$^{+0.49}_{-0.19}$ & - &  0.69 / 96 \\
10 & 535 & APEC & 0.83$^{+0.08}_{-0.13}$ & 0.07$^{+0.05}_{-0.04}$ & - & 0.90 / 87 \\
11 & 1260 & APEC & 0.59$^{+0.03}_{-0.04}$ & 0.19$^{+0.09}_{-0.06}$ & - & 0.68 / 103 \\
12 & 1088 & APEC & 0.71$^{+0.05}_{-0.07}$ & 0.23$^{+0.10}_{-0.07}$ & - & 0.55 / 102 \\
13 & 1080 & APEC & 0.68$^{+0.04}_{-0.05}$ & 0.25 & - & 0.84 / 259 \\
14 & 1279 & APEC & 0.68$^{+0.06}_{-0.03}$ & 0.41$^{+1.21}_{-0.20}$ & - & 0.99 / 272 \\
15 & 1943 & APEC & 0.64$^{+0.05}_{-0.06}$ & 0.25 & - & 1.28 / 373 \\
16 & 2128 & APEC & 0.72$\pm$0.03 & 0.43$^{+0.24}_{-0.13}$ & - & 0.88 / 231 \\
17 & 685 & APEC & 0.67$^{+0.04}_{-0.17}$ & 0.25 & - &  0.92 / 157 \\
\hline
\end{tabular}
\end{center}

\end{table*}

Most of the regions have temperatures of 0.65-0.85~keV, with the
hottest diffuse emission ($>$0.8~keV) located between ESO~514-3 and
NGC~5898. The coolest diffuse emission ($\sim$0.65~keV) is found in the
northern half of the region inside the loop, in the gas in the northwest
parts of NGC~5903, and in the northwest part of the field outside the loop.
The core of NGC~5898 has by far the lowest temperature, but it appears to
have minimal gas content, so this temperature may be affected by
contributions from unresolved spectrally soft point sources in the galaxy. Measured abundances were generally found to be consistent (within 1$\sigma$) with 0.25~\Zsol, except in the region between the loop and NGC~5898, where the best fitting value was 0.07$^{+0.05}_{-0.04}$~\Zsol. Such a low abundance might be taken to indicate the presence of powerlaw emission coming from unresolved point sources, but as the region only partially overlaps the outskirts of NGC~5898, the expected flux from X-ray binaries is small. Additional thermal or powerlaw components do not significantly improve the fit.

The centre panel of Figure~\ref{fig:Tmap} shows a temperature map generated using an adaptive binning technique, as described in \citet{OSullivanetal11a}. A grid of pixels of size 15\arcs\ is established across the field of view, and spectra are extracted from circular regions centred on each pixel size. Extraction region size is allowed to vary in steps of factor 1.4 to achieve a desired number of net counts. In this case 800 net counts were required, producing regions of radius 34-94\arcs. Regions are not independent, leading to an effect similar to adaptive smoothing. Spectra were fitted using an absorbed APEC model. All point sources were removed. The right panel shows the fractional uncertainty on the fitted temperature, and all pixels with uncertainties $>$10 per cent were excluded.

Despite the differences in mapping technique, the basic features of the two maps are similar. Most notably, a line of higher temperatures runs roughly E-W across the lower half of the map, linking ESO~514-3 and NGC~5898, and confirming that this feature is not driven by the region shapes chosen for Figure~\ref{fig:Tmap}. The highest temperatures are located around ESO~514-3, and these may be caused by the inclusion of spectrally hard emission from unresolved X-ray binaries, but we note that Figure~\ref{fig:Tmap} confirms the diffuse emission in this region is hot even if all emission inside the \Dtf\ ellipse of the galaxy is excluded. Temperatures to the north are cooler, but the surface brightness in much of the northwest quadrant is too low for viable spectra to be extracted using the adaptive binning method. The higher temperature region may be associated with the weak X-ray surface brightness enhancement that links ESO~514-3 and NGC~5898 in the \chandra\ image. 

We also extracted ACIS spectra from the low surface brightness channels separating the emission in the core of NGC~5903 from the bright loop sections to its west and south. Spectral fitting did not reveal any significant difference in temperature or hydrogen column when compared to spectra from surrounding regions.

While the structures in the group core are too complex to allow deprojection, we considered it desirable to estimate the properties of the gas within NGC~5903, particularly the cooling time. We used a spectrum from the central 30\arcs\ (4.6~kpc) of the galaxy, using a background taken from a region outside the galaxy to the northeast to subtract off (the majority of) the surrounding group emission. We found the cooling time in this region to be relatively long, 1.70$\pm$0.34~Gyr. This is typical for a non-cool-core group such as NGC~5903 \citep{OSullivanetal17}, and suggests that radiative cooling of the hot gas is unlikely to be a significant source of fuel to the AGN in the current epoch.

\subsection{Point sources in the X-ray ring}
\label{sec:PSs}
We expect to see point sources associated with the group member galaxies, but the X-ray images show a number of sources coincident with the diffuse ring feature. Excluding regions within the \Dtf\ ellipses of group members, and considering the ring to be covered by the elliptical region used in the temperature map, the \textsc{wavdetect} point source search identifies 12 sources in the ring, 8 in the elliptical region enclosed by the ring, and 23 in the remainder of the S3 field of view.

None of the sources have optical counterparts in DSS images, and the SIMBAD database \citep{Wengeretal00} contains no sources within 3\arcs\ of their positions, ruling out foreground objects. We can estimate the number of expected background point sources above a limiting flux from the cumulative number of sources identified by the ChaMP survey \citep{Kimetal07a}. Assuming 10 counts are required for detection, and a powerlaw spectrum with $\Gamma$=1.7 and Galactic absorption, our \chandra\ data are sensitive to sources with flux $F_{0.5-8~keV}\ge$7.23$\times$10$^{-16}$\ergps. We therefore expect to detect $\sim$0.41 sources per square arcminute, or 3.37 background sources in the ring (8.2~arcmin$^2$), 2.94 in the enclosed ellipse (7.16~arcmin$^2$) and 20.4 in the remainder of the field (49.7~arcmin$^2$). While the numbers of sources is small there is some suggestion ($\sim$2$\sigma$ significant) of an excess of point sources in the ring.

The locations of the sources in the ring are not correlated with the \Hi\ structures reported by APW90. None of the sources are correlated with known background objects or with optical sources visible in the DSS. Only two of the sources are bright enough for spectral fitting, and both are consistent with powerlaw emission. Spectral fitting of the diffuse emission in the ring reveals no evidence of a spectrally hard component that might be expected if a significant number of unresolved point sources were present. If we assume the resolved sources to be located within the group, their luminosities cover the range 1-13$\times$10$^{38}$\ergps, consistent with bright X-ray binaries or Ultra-Luminous X-ray sources (ULXs, defined as sources with L$_{0.5-8~keV}\ge$10$^{39}$\ergps). If these were the product of star formation within the ring, a rate of $\sim$0.6-1.3\Msolpyr\ would be needed \citep{Mineoetal12}, with most of the range determined by whether the potential ULXs are included.

\subsection{Active galactic nuclei}
\label{sec:AGN}

We extracted spectra from the central point sources of the three main elliptical galaxies in the \chandra\ S3 field, using 4\arcs\ radius circles in NGC~5903 and NGC~5898, and a 2\arcs\ radius circle for ESO~514-003. All three are well fitted by a powerlaw with Galactic absorption. Table~\ref{tab:AGN} lists their properties. As noted previously, the nucleus of NGC~5903 is detected in the 4.8~GHz data, which suggests the presence of a jet. No radio sources are observed at the positions of the central X-ray sources of NGC~5898 or ESO~514-3. ESO~514-5, the northern spiral galaxy, is outside the field of view of both X-ray observations.

\begin{table}
\caption{\label{tab:AGN}Best-fitting spectral parameters for absorbed powerlaw models fitted to the \chandra\ spectra of the X-ray AGN. Luminosities assume the group distance of 31.5~Mpc.}
\begin{center}
\begin{tabular}{lccc}
\hline
Galaxy & $\Gamma$ & L$_{0.5-7~keV}$ & red. $\chi^2$/d.o.f.\\
 & & (10$^{39}$\ergps) & \\
\hline
NGC~5903 & 2.28$^{+0.24}_{-0.23}$ & 1.49$\pm$0.12 & 0.824/6 \\
NGC~5898 & 1.79$^{+0.09}_{-0.08}$ & 6.74$\pm$0.28 & 0.652/30 \\
ESO~514-003 & 1.48$^{+0.29}_{-0.27}$ & 1.20$\pm$0.15 & 0.929/3 \\
\hline
\end{tabular}
\end{center}
\end{table}

\section{Discussion}
\label{sec:disc}
The combination of X-ray, radio continuum and neutral hydrogen structures in the NGC~5903 group is extremely unusual, with similarities to both AGN lobe/cavity systems, and the collisional shock observed in Stephan's Quintet. We explore the origin of these structures and their relation to the group-member galaxies below.

\subsection{Origin of the IGM structures}
\subsubsection{AGN outburst}
The diffuse radio structure we observe in our GMRT images (and the NVSS image), when combined with the X-ray loop, bears a strong resemblance to a radio lobe and cavity, inflated by the jets of the AGN in NGC~5903. If this is the case, the X-ray loop would be interpreted as a shell of compressed gas enveloping the lobe. The lack of detected polarization and the steep spectral index are both consistent with an old AGN outburst, since we expect the polarization of old radio lobes to be weak, and depolarization by thermal plasma along the line of sight may reduce the polarization further.  The current nuclear activity appears to be low-power, with only kiloparsec-scale jets and no significant re-energization of the large-scale radio structure. NGC~5903 itself shows signs of past disturbance and (possibly recent) episodes of star formation, raising the possibility that the event which triggered the AGN outburst may also have affected the stellar population.

\paragraph{Age of the diffuse radio structure}
We can estimate the age of the diffuse radio structure based on its radio spectrum and luminosity. We note that the following discussion applies to the diffuse structure only, not to the small-scale jet in the core of NGC~5903 nor to the double source near ESO~514-3, whose properties are not sufficiently well defined for such estimates to be made.

Under the assumptions that the relativistic particles and magnetic field are spread evenly throughout the diffuse radio source, and that they are in energy equipartition, we can estimate the magnetic field of the source. We adopt a low energy cutoff of $\gamma_{min}=100$ in the energy distribution of electrons, and a high energy cutoff equivalent to 100~GHz. We approximate the volume of the source as an ellipsoid with semi-major axis 30.5~kpc and semi-minor axis 16~kpc, based on the 234~MHz image. We use a typical spectral index $\alpha$=1, and assume equal energy in electrons and non-radiating particles within the plasma. For these values, we find a minimum energy magnetic field B$_{\rm min}$=3.0~$\mu$G. If the volume were increased by a factor 2 (e.g., if the radio structure is twice as deep as it is wide) B$_{\rm min}$ would be reduced to 2.7~$\mu$G.

Under the assumptions that radiative losses dominate over
expansion losses, that the magnetic field is constant across the source,
and that reacceleration processes have not been important, our observation
that the break frequency of the radio spectrum $\nu_{\rm break}<150$~MHz
can be used to place a lower limit on the age of the source
\citep[e.g.,][]{MyersSpangler85}. The radiative age can be estimated as

\begin{equation}
t_{\rm rad} = 1590  \frac{B_{\rm min}^{0.5}}{(B_{\rm min}^2 + B_{\rm CMB}^2)} [(1+z) \nu_{\rm break}]^{-0.5} \,\,\,\,\,\,\textrm{Myr},
\end{equation}

\noindent where $\nu_{\rm break}$ is expressed in GHz, and $B_{\rm min}$
and $B_{\rm CMB}$ in $\mu$G \citep{Parmaetal07}. $B_{\rm min}$ is the
minimum energy magnetic field, and $B_{\rm CMB} = 3.2(1+z)^2$ is the
equivalent magnetic field of the cosmic microwave background (CMB)
radiation, i.e., the magnetic field strength with energy density equal to
that of the CMB at the redshift $z$.

For B$_{\rm min}$=3~$\mu$G we estimate the lower limit on radiative age to be
$t_{\rm rad} >$360~Myr. This is longer than the age of the \Hi\ filament,
$\sim$200-300~Myr.  The \Hi\ age is estimated from dynamical arguments, the
time required for the filament to extend to its current length given its
internal velocity gradient, and the limit on how long the northern and
southern parts of the filament can remain roughly straight before orbital
motions produce a ``wrapping up'' of the filament around NGC~5903 (APW90).
The filament age is therefore somewhat uncertain, but is probably best
considered as an upper limit. This suggests that the \Hi\ filament is
unlikely to be the source of fuel for the AGN outburst. However, the two
age estimates could be brought into agreement if the radio plasma were out
of equipartition, with an enhanced magnetic field strength.

\paragraph{Outburst energy}
We can estimate the energy released by the AGN outburst from the enthalpy of the radio lobe, 4$pV$ where $V$ is the volume of the cavity and $p$ is the external pressure. The cavity volume is estimated from the inner boundary of the X-ray loop, adopting a prolate ellipsoidal shape for the lobe with semi-major axis 16~kpc and semi-minor axis 13~kpc. The external pressure is taken from the deprojection analysis. The inner edge of deprojected pressure profile is coincident with the outer edge of the X-ray loop at its southern boundary; the surrounding pressure on the northern boundary could be higher owing to the higher IGM density around NGC~5903. 

The external pressure is found to be [7.1$\pm$0.3]$\times$10$^{-13}$~erg~cm$^{-3}$, suggesting a total enthalpy of $\sim$9.2$\times$10$^{56}$~erg. If a second lobe is present, the total for the system would be a factor $\sim$2 greater. We can estimate the jet power required to inflate such a cavity, based on the radiative age derived above, as 8.1$\times$10$^{40}$\ergps. This value is similar to those found in other groups \citep{OSullivanetal11b,Dunnetal10,Cavagnoloetal10,Birzanetal08}.

\paragraph{Triggering mechanism and motion of NGC~5903}
APW90 noted that the radio structure around NGC~5903 resembles a head-tail
source. Our observations confirm that
no second lobe or cavity is visible to the north of NGC~5903. This could be
explained if the galaxy is in motion to the the northeast, with jets
aligned close to the line of sight and the lobes trailing behind the
galaxy, superposed along the line of sight so that they appear as a single
structure. If the galaxy has a velocity component toward us, the more
distant lobe might be beyond the galaxy, explaining some of the radio
emission north and east of the galaxy core, and making its shell of
compressed IGM plasma impossible to distinguish from the overlying X-ray
emission in the galaxy.

The interaction which produced the \Hi\ filament may have been the trigger for the AGN outburst. A tidal interaction might have driven gas in NGC~5903 into the galaxy core, triggering the outburst before stripped \Hi\ entered the galaxy. The age of the radio emission suggests that the \Hi\ is unlikely to have been the fuel source for the outburst, and the \Hi\ disc noted by APW90 looks unlikely as a fueling mechanism for the current activity. The radio jet lies close to the plane of the \Hi\ disc, rather than perpendicular to it. The angle of the disc plane is $\sim$70\degree\ (anti-clockwise from north), the jet axis $\sim$55\degree. By contrast, the H$\alpha$ emission in the galaxy core is aligned with the major axis \citep[i.e.,perpendicular to both the disc and jet,][]{Macchettoetal96}. This suggests that the ionized gas responsible for the H$\alpha$ emission is the more likely fuel source for the AGN, unless the \Hi\ disc undergoes a dramatic alignment change in its inner regions.

\paragraph{Origin of the X-ray loop}
The shell of enhanced X-ray emission surrounding the lobe could be formed in two ways. Either the lobe lifts gas out of the galaxy core as it expands, or if it expands trans-sonically it could compress and shock a shell of IGM gas around it. Uplifted material is likely to retain the temperature and abundance of the gas in the galaxy. Compressed IGM gas should be hotter than the IGM temperature. Unfortunately our observations are not entirely consistent with either scenario. The northern parts of the X-ray loop have temperature similar to the gas in NGC~5903 and the IGM, whereas the southern parts are hotter, but appear to be part of a broader hot feature. Abundances are consistent between the galaxy, IGM and loop. We can estimate the mass of hot gas in the shell from the X-ray data, $\sim$1.5$\times$10$^9$\Msol\ for a $\sim$4.5~kpc thick shell.

The age of the system argues against a shock-heated shell. The spectral index suggests that if we are observing a radio lobe, energy injection from the AGN ceased $>$360~Myr ago. The lobe should by now be close to pressure equilibrium with its surroundings, and any shell should have dispersed back into the IGM. We can estimate the time required for a compressed shell to expand and diffuse into the surrounding IGM after expansion ceases, based on its sound crossing time. For a shell thickness 4.5~kpc and temperature 0.85~keV, we would expect a crossing time of $\sim$12.1~Myr, more than an order of magnitude shorter than the radiative age estimate. For the radiative age to be brought into agreement with the sound crossing time of the shell would require an unrealistically high magnetic field strength $B_{\rm min}$=48~$\mu$G. The presence of a bright cavity rim is therefore in conflict with the apparent age of the radio source.

\subsubsection{Collisional shock}
\label{sec:shock}
Another possibility is that the group is a second example of the type of shock observed in Stephan's Quintet, with a high-velocity collision between a galaxy and the \Hi\ filament producing both X-ray and radio emission. As in the AGN scenario, the filament is formed via a tidal interaction between NGC~5903 and a gas-rich donor. However, in this case the donor loops around NGC~5903 rather than simply passing by the galaxy. This would leave a loop of dense \Hi\ south of NGC~5903, with low density \Hi\ in a diffuse cloud around the galaxy and loop. Another galaxy, passing through the group at high velocity, would collide with these structures, shock heating the southern part of the loop and much of the diffuse \Hi\ to X-ray temperatures.

The radio emission would in this case be a product of the shock. We would expect the pre-shock material to be highly multi-phase, having been stripped from a galaxy, and it would likely contain a population of relativistic electrons (originally produced by supernovae) and ``frozen-in'' magnetic field. As in Stephan's Quintet, the passage of a strong shock through such material would naturally lead to synchrotron emission. The X-ray emission from shocked material would be most visible in the dense loop, but the radio emission would trace both the loop and the low density material. The surface brightness of the radio emission would decrease, and the spectral index increase, as the shock ages. This might explain the low surface brightness ultra-steep 150~MHz emission reported by GK12 east of NGC~5903 --- it would represent the earliest phase of the collision, with the brighter regions being the last parts of the \Hi\ structure to be shock heated.

\begin{figure}
\includegraphics[width=\columnwidth,viewport=20 180 590 710]{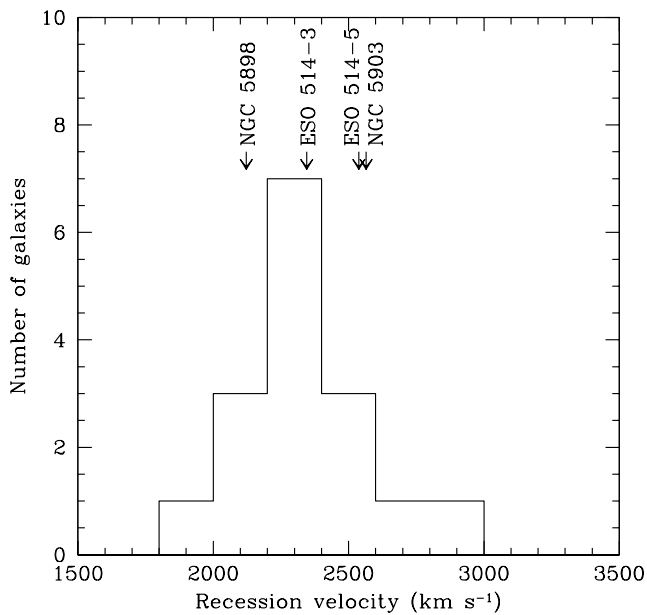}
\caption{\label{fig:gal_vel}Histogram of galaxy velocities for members of the LGG~398 group, using 200\kmps\ bins. The velocities of the major galaxies are indicated by labelled arrows.}
\end{figure}

Since this suggests a galaxy moving from east to west, the obvious candidate for the colliding galaxy is NGC~5898. Its line-of-sight velocity is $\sim$460\kmps\ less than that of NGC~5903 (see Figure~\ref{fig:gal_vel}). We can place a lower limit on its plane-of-sky velocity based on the \Hi\ dynamical timescale. NGC~5898 would need to travel from the eastern edge of the \Hi\ structure to its current location ($\sim$60~kpc) in $<$300~Myr, suggesting a plane-of-sky velocity of $>$200\kmps. If NGC~5903 is approximately stationary with respect to the IGM, this suggests that NGC~5898 has a total relative velocity $>$500\kmps. This would make its motion trans-sonic in the IGM, with Mach number $\mathcal{M}\sim$1.5. Unfortunately both the line-of-sight velocity relative to the IGM and the plane-of-sky velocity are highly uncertain.

\paragraph{Spectral index of shocked emission}
The injection spectral index $\alpha_{inj}$ of shock emission is related to the Mach number as

\begin{equation}
\alpha_{inj} = \frac{\mathcal{M}^2+1}{\mathcal{M}^2-1} - \frac{1}{2},
\end{equation} 

where $\mathcal{M}$ is the Mach number \citep{BlandfordEichler87}. We do not know the injection spectral index, but we can place a lower limit on the Mach number based on the observed spectral index of the diffuse emission. For $\alpha$=1.03, we would require a shock of $\mathcal{M}$=2.18. This implies a shock velocity $\sim$710\kmps, and the shock would heat the 0.65~keV IGM to $\sim$1.5~keV. This is considerably hotter than any temperature currently observed, though post-shock cooling would certainly have an effect. If the injection spectrum was in fact steeper, a weaker shock would be acceptable. The index $\alpha>$1.5 implied for the low-surface brightness emission detected by GK12 would suggest a shock of $\mathcal{M}$=1.73 and a post-shock temperature $\sim$1.1~keV.

The faint trail of X-ray emission linked to NGC~5898 might mark the the path NGC~5898 travelled, where the shock heating of both \Hi\ and IGM was strongest. If the 0.85~keV temperatures observed in the southern parts of the X-ray structure are the result of shock heating in the IGM, they would suggest a shock of $\mathcal{M}$$\sim$1.3, consistent within uncertainties with our initial velocity estimate, but requiring significant cooling to be consistent with the radio spectral index.

\paragraph{Shock heating in the HI filament}
We can calculate the mass of \Hi\ required to form the X-ray loop. Assuming a cylindrical cross-section for the loop, and using the normalization of the X-ray emission in the SW quadrant of the loop, we find a mass in that quadrant of $\sim$7.7$\times$10$^7$\Msol. Taking into account the variation in normalizations around the loop, total shock heated gas mass in the whole structure would be at least 9.4 times larger, so $\sim$7.3$\times$10$^8$\Msol. This would only add about 25\% to the mass of \Hi\ required from the donor galaxy, a plausible amount. 

NGC~5898 would have been highly hypersonic in the cold neutral hydrogen, driving a strong shock. The temperature jump for such a shock would be:

\begin{equation}
\Delta T = \frac{(5\mathcal{M}^2 - 1)(\mathcal{M}^2 +3)}{16\mathcal{M}^2},
\end{equation}

where $\mathcal{M}$ is the Mach number. For an initial temperature of 100K, we would expect a post-shock temperature of $\sim$1~keV. Given that, as in Stephan's Quintet, cooling of the post-shock gas is likely to be initially enhanced by energy losses from dust sputtering and rapid cooling of dense molecular material, this value is probably consistent with the observed temperatures.

\subsubsection{Proposed sequence of interactions -- a combined scenario}
Neither the AGN outburst nor the collisional shock scenario is able to completely explain the observations. The AGN scenario does not explain the correlation between the \Hi\ filament and X-ray loop, the low surface-brightness radio emission observed by GK12, or the high temperatures observed across the southern half of the loop. However, if a shock were the source of the structures, we would expect to see a) the strongest radio emission correlated with the X-ray loop, rather than contained within it, and b) moderate polarization \citep[up to 30\%,][]{Kempneretal04} of the diffuse radio structure, caused by the shock compressing and ordering the magnetic field. The radio emission from the shock in Stephan's Quintet is observed to be polarized \citep{NikielWroczynskietal13}.

A combination of the two scenarios may be more successful. We can hypothesize a series of interactions between the galaxies in the group as follows: i) A few hundred Myr ago, ESO~514-5 (or some other donor) falls through the group, passing close enough to NGC~5903 for tidal stripping to take place, creating the \Hi\ filament and a more diffuse scattering of \Hi\ clouds through the group core; ii) The tidal forces trigger an AGN outburst in NGC~5903, with a jet aligned close to our line of sight inflating radio lobes; iii) The motion of NGC~5903 to the northeast means that the radio lobes are left behind it, producing the northeast-southwest alignment of the bright radio structure; iv) the AGN exhausts its fuel supply, jet power is reduced to the current levels, and the lobes are left to passively age; v) NGC~5903 and the northern part of the radio lobes move up against the \Hi\ filament, whose central parts fall into NGC~5903, forming the \Hi\ disc; vi) Meanwhile, NGC~5898 falls through the group from east to west at high velocity, losing any hot gas in its own halo and leaving a trail of enhanced temperatures behind it; vi) Although NGC~5898 does not directly collide with the \Hi\ filament, collisions with the diffuse \Hi\ component leads to low surface-brightness radio emission and perhaps enhanced X-ray emission across the group core.

In this scenario the X-ray loop would be primarily formed from gas uplifted out of NGC~5903, and the apparent high temperatures in its southern half would be caused by superimposed higher temperature material shock heated by the passage of NGC~5898. The brightest parts of the radio structure would be radio lobes from the outburst in NGC~5903, aligned close to the line of sight so that only a single lobe, cavity and shell is visible. 

This combined scenario moves us closer to a complete explanation, but it requires the recent history of the group to have included a sequence of violent interactions, and its complexity may make it less plausible. It also requires the correlation between the X-ray loop and \Hi\ filament to be largely serendipitous, and does not explain the enhanced X-ray emission where the \Hi\ and loop overlap. While studies of Stephan's Quintet have shown that cold molecular gas and dust can enhance the cooling rate of $\sim$1~keV material \citep{Guillardetal09, Cluveretal10}, there is no evidence of enhanced X-ray luminosity correlated with \Hi\, and our \xmms\ observation of NGC~5903 does not reveal any clear X-ray structure associated with the \Hi\ in the northern part of the filament. Nonetheless, this may be the most comprehensive picture we can draw of the evolution of the group core, within the limits of the available data.

New observations in the optical and IR may be able to provide additional information. The shock in Stephan's Quintet is a strong H$\alpha$ \citep{Sulenticetal01} and mid-IR source \citep{Xuetal99}, and appears to have triggered star formation \citep[e.g.,][]{MendesdeOliveiraetal98,Xuetal05,Fedotovetal11}. No H$\alpha$ imaging of NGC~5903 is available, and IRAS IR imaging lacks the spatial resolution to distinguish the region of interest. Digitized Sky Survey optical imaging shows no obvious star forming regions, but the possible excess of X-ray point sources associated with the ring might be an indicator that at least some star formation has taken place. Follow-up observations of NGC~5903 could therefore help resolve the question of whether a shock has played a significant role in the formation of the structures in the NGC~5903 group.

\subsection{Possible HI donors} 
APW90 discuss the possible identity of the galaxy responsible for the \Hi\ filament, and come to the conclusion that ESO~514-5 is the most likely candidate. We will briefly revisit this question in the light of our new data.

ESO~514-3, the galaxy immediately south of NGC~5903, has been suggested as a likely donor (e.g., by GK12) owing to its location close to the southern tip of the \Hi\ filament. However, APW90 detect no \Hi\ in the galaxy and note that it is actually located just outside the filament, and that it is an early-type galaxy. ESO~514-3 is most plausible as the \Hi\ donor if the AGN scenario discussed above is true. If the X-ray loop was once part of the \Hi\ filament, it is difficult to see how ESO~514-3 could have looped around NGC~5903 and returned to its current position. 

ESO~514-5, located $\sim$16\arcm\ (168~kpc) north of NGC~5903, is an edge-on Sa galaxy with a prominent dust lane, containing 5$\times$10$^8$\Msol\ of \Hi\ with a centrally-peaked (rather than disc-like) velocity profile \citep{Theureauetal98}. Its axis is roughly aligned toward NGC~5903, and it lies close to the line of extension of the northern \Hi\ filament. Our continuum image suggests at least some ongoing star formation in the galaxy disc. The galaxy properties are therefore generally consistent with what might be expected after ram-pressure and/or tidal stripping. 

Taking the velocity dispersion of the group, 400\kmps, as an estimate of the galaxy's plane-of-sky motion, it would take $\sim$410~Myr to travel the distance from NGC~5903. Given the large uncertainties, this is probably consistent with the \Hi\ dynamical timescale. Both the scenarios for formation of the X-ray/radio structures discussed above would be consistent with ESO~514-5 as the \Hi\ donor, with the galaxy either falling into the group from the south and passing straight through, or falling in from the north and looping around NGC~5903. We agree with APW90 that ESO~514-5 is the most likely donor galaxy.

\subsection{The nature of the ESO~514-3 double-lobed radio source}
\begin{figure}
\includegraphics[width=\columnwidth,viewport=45 75 560 720]{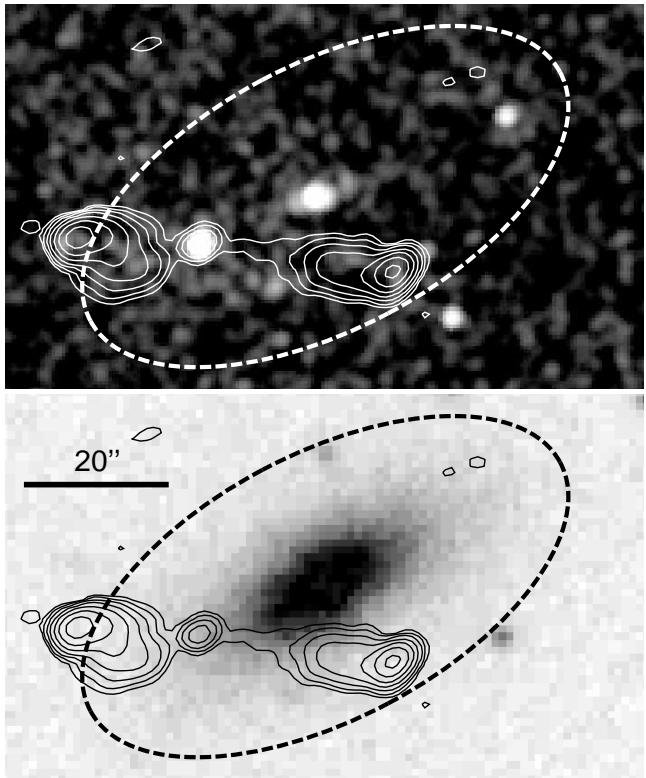}
\caption{\label{fig:ESO}\chandra\ 0.5-7~keV Gaussian smoothed (\textit{upper}) and DSS optical (\textit{lower}) images of ESO~514-003, overlaid with the \Dtf\ ellipse (dashed line) and VLA 4.8~GHz contours (solid lines). The VLA contours have HPBW=3.6$\times$2.45\arcs\, p.a.=114\degree\, and start at 3$\times$r.m.s=51$\mu$Jy, stepping up by factors of 2. The core of the radio source is coincident with an X-ray point source, and offset from the galaxy centre by $\sim$20\arcs.}
\end{figure}

The question of whether the double radio source NVSS~J151837-240720 is associated with ESO~514-3 has been the subject of some discussion in the literature \citep{GopalKrishna78,Appletonetal90special,GopalKrishnaetal12special}. The new \chandra\ observation and newly analysed VLA C-band data provide a conclusive answer: the radio source is a background object, and is not associated with ESO~514-3.

Figure~\ref{fig:ESO} shows the VLA 4.8~GHz continuum contours overlaid on optical and \chandra\ X-ray images of ESO~514-3. The VLA data show that the core of NVSS~J151837-240720 is located at \mbox{15$^h$18$^m$37\fd 1, -24\degr 07\arcm 18\arcs}, $\sim$17\arcs\ to the east of the galaxy centre, and coincident with an X-ray point source. Spectral fitting (for a 2\arcs\ radius region) shows that this X-ray source is well described by an absorbed powerlaw model with $\Gamma$=1.63$\pm$0.13, and a 0.5-7~keV flux of 2.94$\pm$0.18$\times$10$^{-14}$\ergpspcmsq. Given the positional offset, the radio morphology, the polarization of the western component and the correlation of the core with a spectrally hard X-ray source, it seems clear that this is a background radio galaxy behind ESO~514-3. Unfortunately NED lists no corresponding optical, IR or UV source from which a redshift might be determined.

\section{Summary and Conclusions}
\label{sec:conc}
The NGC~5903/NGC~5898 system is an example of the later stages of group evolution. Two large ellipticals dominate the galaxy population, though both still retain signs of the mergers and interactions which formed them. The group has an extended halo of hot, 0.65~keV X-ray emitting gas, but also contains a 110~kpc long \Hi\ filament, stripped from one of the late-type group members which still retain some cold gas. In short, while the group has completed most of the evolutionary path from spiral-dominated/cold-gas-rich to elliptical-dominated/hot-gas-rich system, some important (and indeed spectacular) interactions are still in progress.

Attention has been drawn to the group by the anti-correlated \Hi\ and radio continuum structures. Our own observations were designed to elucidate the origin of these structures, and we have made significant progress in understanding this unusual group. Our results can be summarized as follows:

\begin{enumerate}
\item We report the first detection of a hot intra-group medium in the group. X-ray emission is detected out to the edge of the \xmms\ field of view ($\sim$950\arcs\ or $\sim$145~kpc). The typical IGM temperature in the region surrounding the dominant ellipticals is $\sim$0.65~keV, and typical abundance is $\sim$0.25~\Zsol. The total 0.3-7~keV gas luminosity in the \xmms\ field of view is 1.63$^{+0.06}_{-0.05}$$\times$10$^{41}$\ergps.

\item Both ellipticals possess central X-ray point sources, as does the nearby small early-type galaxy ESO~514-3. All three are well described by absorbed powerlaw spectra with luminosities of a few 10$^{39}$\ergps, consistent with emission from AGN. The source in NGC~5903 is coincident with a 4.8~GHz radio source whose extension suggests a small-scale jet. 

\item NGC~5903 lies at the peak of the diffuse X-ray emission and is evidently hot-gas-rich. Diffuse X-ray emission is aligned from northwest to southeast along the major axis of the galaxy and continues beyond the stellar population to form a roughly elliptical loop $\sim$35~kpc across, extending to the southwest of the galaxy and overlapping ESO~514-3. The northern, brighter parts of the loop are correlated with southern part of the \Hi\ filament. The area inside the loop is correlated with the brightest part of the diffuse radio structure, though lower surface brightness radio emission extends beyond NGC~5903 to the northeast and beyond the eastern edge of the loop. NGC~5898 and ESO~514-3 are poor in hot gas.

\item We used the GMRT to observe the group at 234, 612 and 1410~MHz. Adding the TGSS 150~MHz data and flux measurements from the literature, we find that the bright diffuse radio emission inside the X-ray loop has a steep spectral index $\alpha_{150}^{612}$=1.03$\pm$0.08. No break is seen in the radio spectrum, suggesting that the diffuse radio structure is $>$360~Myr old (for magnetic field strength B$_{\rm min}$=3.0~$\mu$G). We map the spectral index distribution and find that the flattest indices ($\alpha$$\sim$0.8) are located close to the AGN in NGC~5903 and along the ridge of brightest radio emission extending southwest. Steep indices ($\alpha$$>$1.6) are found at the edge of the diffuse radio emission, and even steeper indices are implied for the low surface brightness emission southeast of NGC~5903 reported by GK12. NGC~5903 hosts a small-scale jet (1.4~kpc long), but while this may contribute to the flatter spectral index in the core of the galaxy, it is clearly too weak to have a significant impact on the larger diffuse structure.

\item Using \chandra\ to map the temperature of the IGM and loop structure, we find a line of enhanced temperatures (kT$\sim$0.85~keV) extending E-W across the southern half of the loop from ESO~513-3 to NGC~5898. This coincides with a faint trail of X-ray emission connecting the loop to NGC~5898. Moderate temperatures ($\sim$0.7~keV) are found in NGC~5903 and the eastern part of the loop, and cooler temperatures ($\sim$0.65~keV) to the northwest.

\item We consider possible origin scenarios for the correlated X-ray loop, \Hi\ filament and (anti-correlated) diffuse radio structure. An AGN outburst could produce a set of radio lobes with shells of uplifted or shocked hot gas, aligned along the line of sight so that only a single lobe/shell is visible. The 4$pV$ enthalpy of such an outburst would have been 1.8$\times$10$^{57}$~erg, with jet powers of order 8$\times$10$^{40}$\ergps, consistent with other group-dominant radio galaxies. The anti-correlation with the \Hi\  might be the product of the expanding lobes pushing up against the filament, but this does not explain the brightness of the cavity rim in areas where it coincides with the \Hi\ filament, nor the origin of the high temperatures in the southern part of the cavity. Alternatively, the X-ray and radio structures might be the result of the partial shock-heating of the \Hi\ loop during a high-velocity collision with NGC~5898. A shock of Mach $\sim$1.3 would be sufficient to produce the high temperatures seen in the south of the group, and this is broadly plausible given the limits on the velocity of NGC~5898 relative to the IGM. Such a shock would also explain the radio emission outside the X-ray loop and particularly the low surface brightness, steep spectral index emission reported by GK12; these would represent the first regions of diffuse \Hi\ to be shocked. However, this scenario also has flaws: we see no optical indicators of such a large-scale shock, the spectral index map does not show the expected E-W gradient, and most importantly X-ray and radio emission are not correlated in the X-ray loop. We consider it likely that a combination of shock and AGN outburst is required to explain the observations, and note that this requires a sequence of violent interactions in the recent history of the group.

\item We report the first high-resolution 1.4~GHz continuum maps of the edge-on spiral galaxy ESO~514-5. The 1.4~GHz map reveals a central point source, likely an AGN, with a narrow extension aligned with the galaxy disc, probably arising from star formation. We agree with APW90 that ESO~514-5 is most likely the donor from which the \Hi\ filament was stripped.

\item Using archival VLA 4.8~GHz data, we show that the bright double radio source NVSS J151837-240720 is located southeast of the nucleus of ESO~514-3. The source is resolved into a typical double-lobed structure with a well-defined point-like core, whose position closely agrees with that of a \chandra\ X-ray point source. The X-ray spectrum of the source is well modelled as an absorbed $\Gamma$$\sim$1.6 powerlaw with 0.5-7~keV flux $\sim$3$\times$10$^{-14}$\ergpspcmsq. The western lobe is shown to be mildly ($\sim$5\%) polarized at 608~MHz. These data resolve the question of the nature of this source; it is clearly a background radio galaxy. 

\end{enumerate} 

Further observations are needed to determine the sequence of events which produced the current disturbed state of the NGC~5903 group. As noted in Section~\ref{sec:shock}, H$\alpha$ imaging could be decisive if it were to reveal ionized gas correlated with the X-ray loop. This would strongly imply formation through shock heating of \Hi, as in the shock in Stephan's Quintet. Deeper neutral hydrogen observations are needed to confirm the link between the \Hi\ filament and ESO~514-5, and to determine the degree of disturbance in that galaxy. More detailed \Hi\ velocity mapping of the filament might provide clues to the nature of its interaction with NGC~5903, the radio structure and X-ray loop. We plan to present deeper GMRT \Hi\ observations of the group in a future paper, and hope to provide a more definitive picture of the interactions in this unusual and interesting system.

\section*{Acknowledgments}
The authors thank the anonymous referee for their thorough reading of the paper and helpful comments. We thank the staff at the GMRT and National Center for Radio Astrophysics (NCRA) for their support, which made our observations possible. The GMRT is a national facility operated by the NCRA of the Tata Institute for Fundamental Research (TIFR). The scientific results reported in this article are based to a significant degree on observations made with the \chandra\ and \xmm\ X-ray Observatories. Support for this work was provided by the National Aeronautics and Space Administration through XMM-Newton award number NNX14AF04G, and through Chandra Award Number G05-16123X issued by the Chandra X-ray Observatory Center, which is operated by the Smithsonian Astrophysical Observatory for and on behalf of the National Aeronautics Space Administration under contract NAS8-03060. This research has made use of the SIMBAD database, operated at CDS, Strasbourg, France.

\bibliographystyle{mnras}
\bibliography{../paper}

\label{lastpage}
\end{document}